\documentclass[a4paper,11pt,preprintnumbers]{article}
\usepackage{jcappub} 
\usepackage{mathtools}
\usepackage{booktabs}
\usepackage[english]{babel}
\usepackage{amsmath,amssymb,amsbsy,amstext, amsthm, simplewick, amsfonts}
\usepackage{hyperref}
\usepackage{graphicx}
\usepackage[small]{caption}
\usepackage{siunitx}
\usepackage{upgreek}
\usepackage{framed}
\usepackage{wrapfig}
\usepackage{multirow}
\usepackage{bbm}
\usepackage[svgnames,dvipsnames,x11names]{xcolor}
\usepackage[utf8x]{inputenc}
\usepackage{selinput}
\usepackage{bm}
\usepackage{float}
\usepackage{geometry}
\usepackage{yfonts}
\usepackage{caption}
\usepackage{subcaption}
\usepackage{subfloat}
\usepackage{sidecap}
\usepackage{longtable}
\usepackage{physics}    
\usepackage{anyfontsize}
\usepackage{dsfont}
\usepackage{colortbl}
\usepackage{lineno}
\usepackage{enumitem}
\usepackage{comment}
\usepackage{fancyhdr}

\newcommand{\rom}[1]{\uppercase\expandafter{\romannumeral #1\relax}}

 \arxivnumber{YITP-24-25} 
\title{Enhanced Curvature Perturbation and Primordial Black Hole Formation in Two-stage Inflation with a break}

\author{Xinpeng Wang$^{a,d}$,Ying-li Zhang$^{a,b,c,d,e}$, Misao Sasaki$^{d,f,g}$}

\affiliation{$^a$ School of Physics Science and Engineering, Tongji University, Shanghai 200092, China\\
$^b$Institute for Advanced Study of Tongji University, Shanghai 200092, China\\
$^c$Institute of Theoretical Physics, 
Chinese Academy of Sciences, Beijing 100190, China 
$^d$Kavli Institute for the Physics and Mathematics of the Universe (WPI), The University of Tokyo Institutes for Advanced Study, The University of Tokyo, Chiba 277-8583, Japan\\
$^e$Center for Gravitation and Cosmology, Yangzhou University, Yangzhou 225009, China\\
$^f$Center for Gravitational Physics and Quantum Information,
Yukawa Institute for Theoretical Physics, Kyoto University, Kyoto 606-8502, Japan\\
$^g$Leung Center for Cosmology and Particle Astrophysics,
National Taiwan University, Taipei 10617, Taiwan}

\emailAdd{xinpengwang@tongji.edu.cn, yingli@tongji.edu.cn, misao.sasaki@ipmu.jp}

\abstract{
We investigate a model of $R^2$-gravity with a non-minimally coupled scalar field that gives rise to two-stage inflation with a break, that is, with an intermediate stage where inflation momentarily halts. 
We find that the power spectrum of the primordial curvature perturbation is significantly enhanced at the break scale, which can account for the primordial black hole (PBH) formation,
without affecting the CMB constraint on large scales. 
The behavior of the curvature perturbation is carefully analyzed and we find a few notable new features in the spectrum. In particular, we find that the $k^3$ growth of the spectrum of toward the end of the first stage of inflation. We argue that this is a universal feature common to all two-stage models where the field dominating the second stage is heavy during the first stage.
By appropriately tuning the model parameters, we find that our model can realize the scenario of PBHs as the cold dark matter of the Universe.
We also find that we can choose the parameters so that the spectrum of the induced gravitational waves from the enhanced curvature perturbation fits the NANOGrav-15yr data of pulsar timing array observation.
}

\begin{document}

\maketitle
\flushbottom
\section{Introduction}

Inflation is now an integrated part of the standard cosmological model, as it can provide the initial seeds for
all the structures in the Universe, including life. On large scales, say $\gtrsim 1\,{\rm Mpc}$, the observed fluctuations in the cosmic microwave background radiation (CMB) and the large-scale structure (LSS) are perfectly consistent with typical models of inflation, and the primordial curvature perturbation amplitude is constrained to be $\sim10^{-9}$ at high accuracy~\cite{Planck:2018jri}. These highly constrained scales correspond to the comoving scales that left the Hubble horizon at the number of $e$-folds $N\simeq40\sim60$ back from the end of inflation in typical models. 
However, on much smaller scales, say corresponding to $N\lesssim30$, there are virtually no observational data that can constrain models of inflation, due to various nonlinear astrophysical processes that destroyed the memories of the initial condition during the history of the Universe.

This situation has started to change in recent years thanks to the rapid progress in gravitational wave cosmology.
Gravitational waves (GWs) that may be produced in the early Universe, either during or after inflation, may reach us without the initial condition memories being destroyed, as their interaction with matter is extremely weak.
As one of such possibilities, the GWs generated at second order from an enhanced primordial curvature perturbation, the so-called scalar-induced GWs (SIGWs), have become an important observational target~~\cite{Matarrese:1992rp, Matarrese:1993zf, Nakamura:1996da, Matarrese:1997ay, Noh:2004bc, Carbone:2004iv, Nakamura:2004rm,Ananda:2006af, Osano:2006ew}.

An intriguing effect associated with such an enhanced curvature perturbation is the formation of black holes.
Even if the mean amplitude of the curvature perturbation is not extremely large, the probability of realizing a rare but large amplitude perturbation is non-vanishing due to its stochastic nature. If the spatial curvature of such a region is large and positive when its scale re-enters the Hubble horizon after inflation, that region will collapse to form a black hole. They are called primordial black holes (PBHs), and their existence was suggested more than a half-century ago~\cite{zel1966hypothesis, Hawking:1971ei, Carr:1974nx, Meszaros:1974tb, Khlopov:1985jw}.

PBHs contain rich phenomena of physics. First, since PBHs interact with other matter only via gravity, and they are born static, they are a natural candidate for cold dark matter (CDM)~\cite{Chapline:1975ojl}.
In particular, for asteroid mass (ranging from $10^{-16}M_\odot$ to $10^{-12}M_\odot$) PBHs,
they may constitute all the CDM~\cite{Niikura:2019kqi, Katz:2018zrn, Smyth:2019whb, Montero-Camacho:2019jte}. 
Second, PBHs also provide a natural explanation for the high masses and low spins of binary black hole (BBH) ~\cite{Bird:2016dcv, Clesse:2016vqa, Sasaki:2016jop,Somiya:2011np,Aso:2013eba,Wang:2022nml,Franciolini:2022tfm,Kimura:2021sqz,Franciolini:2023opt}, inferred from the past three runs of GW observations by LIGO-Virgo and LIGO-Virgo-KAGRA collaborations~\cite{LIGOScientific:2018mvr, LIGOScientific:2020ibl, LIGOScientific:2021djp}. 
The recently discovered signal of a GW background by several pulsar timing array (PTA) collaborations at $10^{-9}\sim10^{-8}$ Herz band ~\cite{NANOGrav:2023hvm, Reardon:2023gzh, Antoniadis:2023rey, Xu:2023wog}  has drawn a widespread attention \cite{, Balaji:2023ehk, You:2023rmn, Yi:2023mbm, Cai:2019elf, Domenech:2024rks,Cai:2023dls,Liu:2023pau,Yi:2023npi,Liu:2023hpw,Chen:2024fir,Chen:2024twp},  it indicates the existence of sub-solar mass PBHs, which may be detected by microlensing observations \cite{SupernovaCosmologyProject:1993faz, Aubourg:1993wb, MACHO:2000qbb, POINT-AGAPE:2005swi, Dong:2007px, Wyrzykowski:2010mh, Wyrzykowski:2011tr, Niikura:2017zjd, Niikura:2019kqi}.

There are a number of proposed mechanisms to enhance the primordial curvature perturbations during inflation, for instance, single-field inflation with a non-slow-roll stage ~\cite{Leach:2001zf,Byrnes:2018txb,Domenech:2023dxx,Ozsoy:2019lyy,Pi:2022zxs,Cai:2022erk,Ragavendra:2020sop,Fu:2020lob,Vennin:2020kng,Atal:2019erb,Liu:2020oqe,Bhaumik:2019tvl,Mishra:2019pzq,Dalianis:2018frf,Garcia-Bellido:2017mdw,Garcia-Bellido:2016dkw,Yokoyama:1998pt,Ivanov:1994pa,Starobinsky:1992ts,Di:2017ndc,Zhai:2022mpi,Balaji:2022rsy,Choudhury:2024one,Choudhury:2023hfm,Caravano:2024tlp,Heydari:2023rmq,Heydari:2021gea,Cai:2023uhc}, multi-field inflation~\cite{Sasaki:1998ug,Garcia-Bellido:1996mdl,Kawasaki:1997ju,Frampton:2010sw,Giovannini:2010tk,Clesse:2015wea,Inomata:2017okj,Gordon:2000hv,Anguelova:2020nzl,Bhattacharya:2022fze,Braglia:2020eai,Braglia:2020fms,Ema:2020zvg,Cotner:2019ykd,Gundhi:2018wyz,Christodoulidis:2023eiw,Inomata:2018cht,Fumagalli:2020adf,Cai:2021wzd,Meng:2022low,Dimastrogiovanni:2024xvc,Qin:2023lgo}, modify gravity~\cite{Karam:2018mft,Enckell:2018uic,Samart:2018mow,Fu:2019ttf,Gundhi:2020kzm,Frolovsky:2022ewg, Aldabergenov:2020bpt,Karydas:2021wmx,Lee:2021rzy,Canko:2019mud,Pi:2017gih,He:2020qcb,Cheong:2020rao,Cheong:2019vzl,He:2020ivk,He:2018gyf,Cado:2023zbm,Kawai:2022emp,Cheong:2022gfc,Chen:2021nio,Kawai:2021edk,Zhang:2021rqs,Yi:2022anu,Qiu:2022klm,Lin:2020goi,Yi:2020cut,Yi:2020kmq,Gao:2021vxb,Lin:2021vwc,Chen:2024gqn}, curvaton scenario~\cite{Pi:2021dft,Lyth:2002my,Kawasaki:2012wr,Kohri:2012yw,Ando:2017veq,Ando:2018nge}, 
parametric resonance~\cite{Cai:2018tuh,Chen:2020uhe,Chen:2019zza,Cai:2019bmk}, etc. The property that any successful model must have is its consistency with the CMB and LSS constraints on large scales.
In other words, it is essential to decouple the physics on small scales from that on large scales. 
This can be naturally realized in the scenario where inflation consists of two or more distinct periods intervened by a transition stage that may or may not break inflation. We call it the two-stage inflation. It was proposed by various people for various purposes~\cite{Kofman:1985aw, Silk:1986vc, Zelnikov:1991nv, Polarski:1992dq, Polarski:1995zn,Balaji:2022dbi}.

In this paper, we investigate the enhancement of primordial curvature perturbations and the PBH formation in the two-stage inflationary scenario. As a concrete model, we adopt Starobinsky's $R^2$ gravity theory first proposed in~\cite{Starobinsky:1980te} supplemented with a non-minimally coupled scalar field, where the first stage is dominated by the scalaron of the $R^2$ theory while the second stage by the non-minimally coupled scalar field. 
The main reason for this choice is that it is easy to satisfy the observational constraint on the CMB scales while giving us enough degrees of freedom to design the second stage.

The paper is organized as follows: In Section \ref{model}, we introduce the Lagrangian of our model and overview its features. 
In Section~\ref{background}, We solve the background evolution of our model. In order to do so, we
divide the whole process into three stages, the first inflationary stage, the second intermissionary stage, and the last inflationary stage. 
In Section~\ref{perturbation}, we study the curvature perturbation in linear theory. We calculate the power spectrum of the primordial curvature perturbation by numerically solving the perturbation equations. We also derive an approximate analytical formula for the curvature perturbation by using the $\delta N$ formalism.
In Section~\ref{appliance}, we consider the PBH formation and the SIGWs in our model. For simplicity, we adopt the simple Press-Schechter criterion for the PBH formation. We find that the PBH mass function is sharply peaked. 
As an example, we apply our model to several different mass ranges of PBHs and discuss the implications. In particular, in the case of sub-solar mass PBHs, the resulting SIGWs seem to fit the NANOGrav-15yr data~\cite{NANOGrav:2023hvm} fairly well.
The conclusion and discussion are given in Section~\ref{conclusion}.  
In Appendices, we summarize the description of multi-field inflation including the background evolution and the field perturbation evolution for our model. 

\section{The model}
\label{model}
We consider the action where $R^2$ gravity is non-minimally coupled by a scalar field $\chi$~\cite{Pi:2017gih},
\begin{align}\label{Jordan}
    S_{J}=\int d^4x\sqrt{-g}\ \bigg[\frac{M_{\mathrm{pl}}^2}{2}f(R,\chi)-\frac{1}{2}g^{\mu\nu}\partial_\mu\chi\partial_\nu\chi-V(\chi)\bigg],
\end{align}
where $M_{\mathrm{pl}}\equiv(8\pi G)^{-1/2}$, $f(R,\chi)$ is given by 
\begin{align}\label{fdef}
    &f(R,\chi)=R+\frac{R^2}{6M^2}-\frac{\xi R}{M_{\mathrm{pl}}^2}(\chi-\chi_0)^2,
\end{align}
and $V(\chi)$ is a Higgs-like potential,
\begin{align}\label{Vchi}
    &V(\chi)= V_0-\frac{1}{2}m^2\chi^2+\frac{1}{4}\lambda\chi^4.
\end{align}
In the above, $\chi_0$ is a constant to break the $Z_2$ symmetry, $\chi\leftrightarrow-\chi$, in order to avoid too large quantum fluctuations at the second inflationary stage (to be explained later), and $V_0$ may be expressed in terms of $m$ and $\lambda$ as
\begin{align}\label{V0mL}
    V_0=\frac{m^4}{4\lambda}\,,
\end{align}
to guarantee  the vanishing of the cosmological constant at the end of inflation.

Under the conformal transformation $\tilde g_{\mu\nu}=F g_{\mu\nu}$ with
\begin{align}
    &F\equiv\partial f/\partial R=1+\frac{R}{3M^2}-\xi\left(\frac{\chi-\chi_0}{M_{\mathrm{pl}}}\right)^2,
\end{align}
the action \eqref{Jordan} can be expressed in the Einstein frame as 
\cite{Dicke:1961gz,Faraoni:1998qx,Fujii:2003pa,DeFelice:2010aj,Nojiri:2010wj,Nojiri:2017ncd,Hell:2023mph}
\begin{align}\label{EinsteinS}
    S_{\mathrm{E}}=\int d^4 x\sqrt{-\tilde g}\left[\frac{M_{\mathrm{pl}}^2}{2}\tilde R-\frac{1}{2}\tilde g^{\mu\nu}\partial_{\mu}\phi\partial_{\nu}\phi-\frac{1}{2}F^{-1}\tilde g^{\mu\nu}\partial_\mu\chi\partial_\nu\chi-U(\phi,\chi)\right],
\end{align}
where the scalar field $\phi$ (called the scalaron) and the potential $U(\phi,\chi)$ are defined as
\begin{align}    
\phi&\equiv M_{\mathrm{pl}}\sqrt\frac{3}{2}\ln F\,,\label{phidef}\\
U(\phi,\chi)&
\equiv\frac{3}{4}M^2M_{\mathrm{pl}}^2W^2(\phi,\chi)+\frac{V(\chi)}{F(\phi)^2}\,,\label{Udef}
\end{align}
and we have introduced a dimensionless function $W(\phi,\chi)$ defined by
\begin{align}\label{Wdef}
W(\phi,\chi)\equiv 1-\frac{1}{F(\phi)}\left[1-\xi\left(\frac{\chi-\chi_0}{M_{\mathrm{pl}}}\right)^2\right].
\end{align}

\begin{figure}[htbp]
\centering
\includegraphics[width=0.7\textwidth]{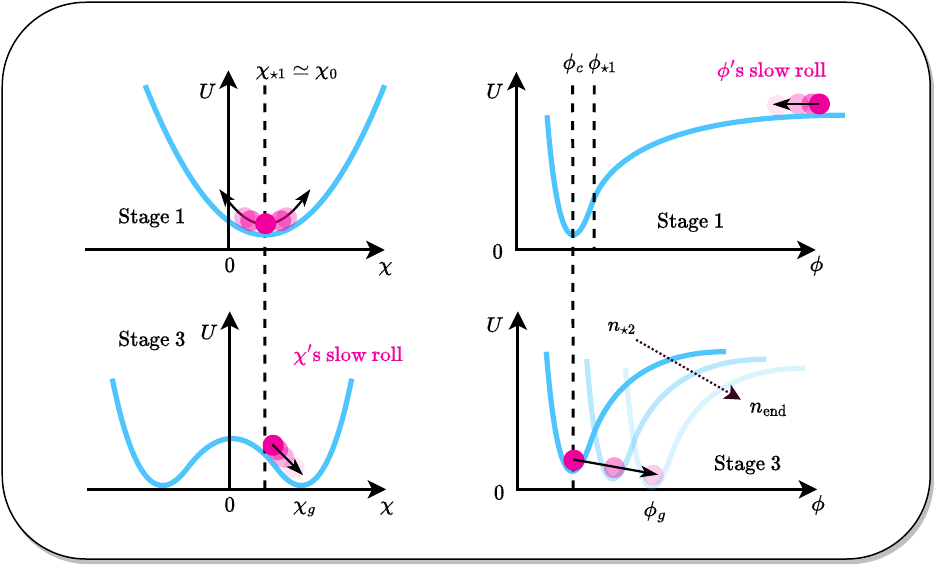}
\caption{A schematic diagram of the potential.
The upper panels are for St-1. The upper-left panel shows a concave potential that traps $\chi$, and the upper-right panel shows slow-roll inflation dominated by $\phi$ on the potential plateau along $\phi$ direction.
The lower panels are for St-3.
The lower-left panel shows a Mexican hat potential which achieves slow-roll inflation dominated by $\chi$ during St-3,
while the lower-right panel shows the effective potential that traps $\phi$ in the potential valley.}
\label{potential}
\end{figure}

\begin{figure}[htbp]
\centering
\includegraphics[width=0.7\textwidth]{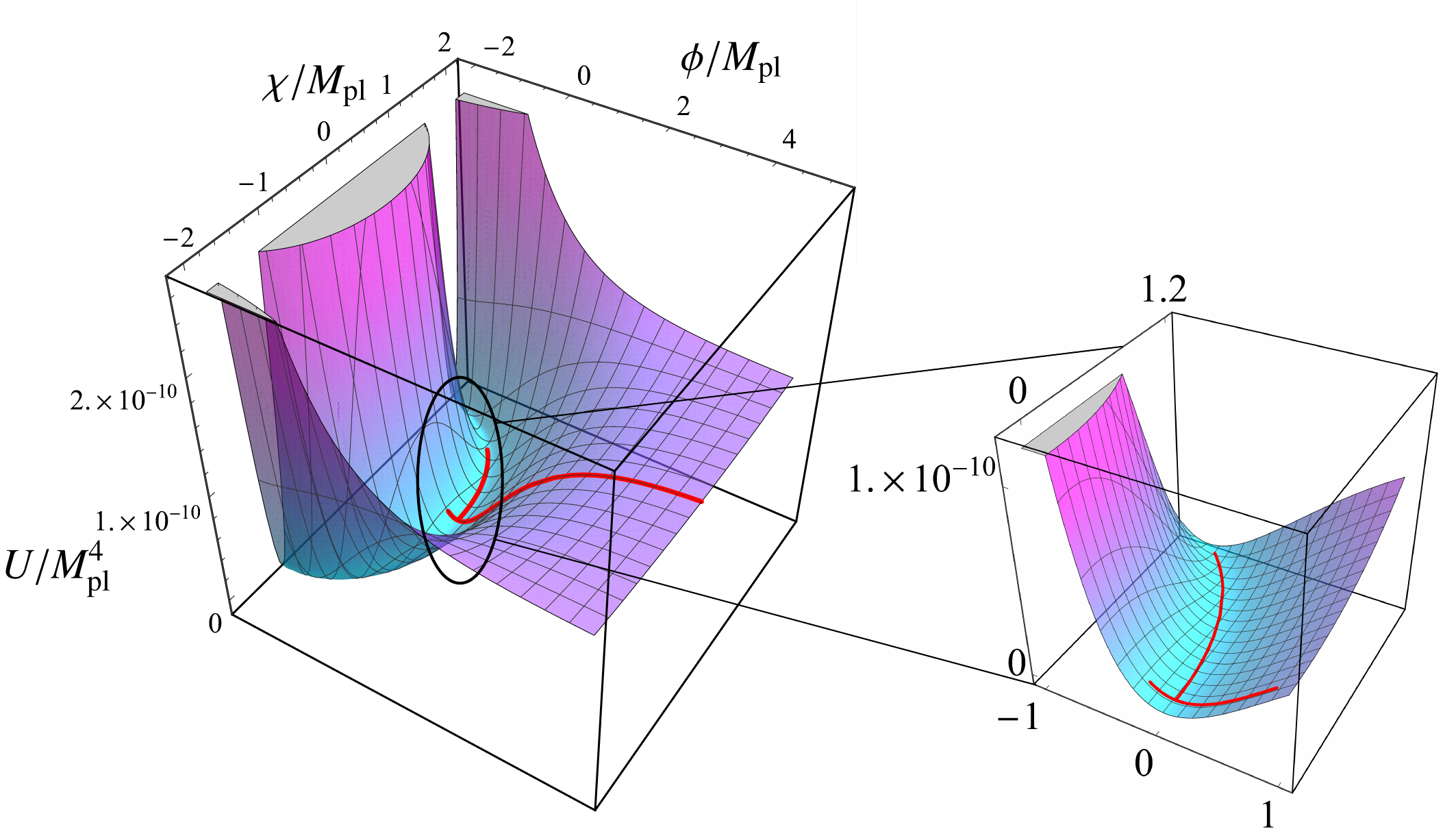}
\includegraphics[width=0.4\textwidth]{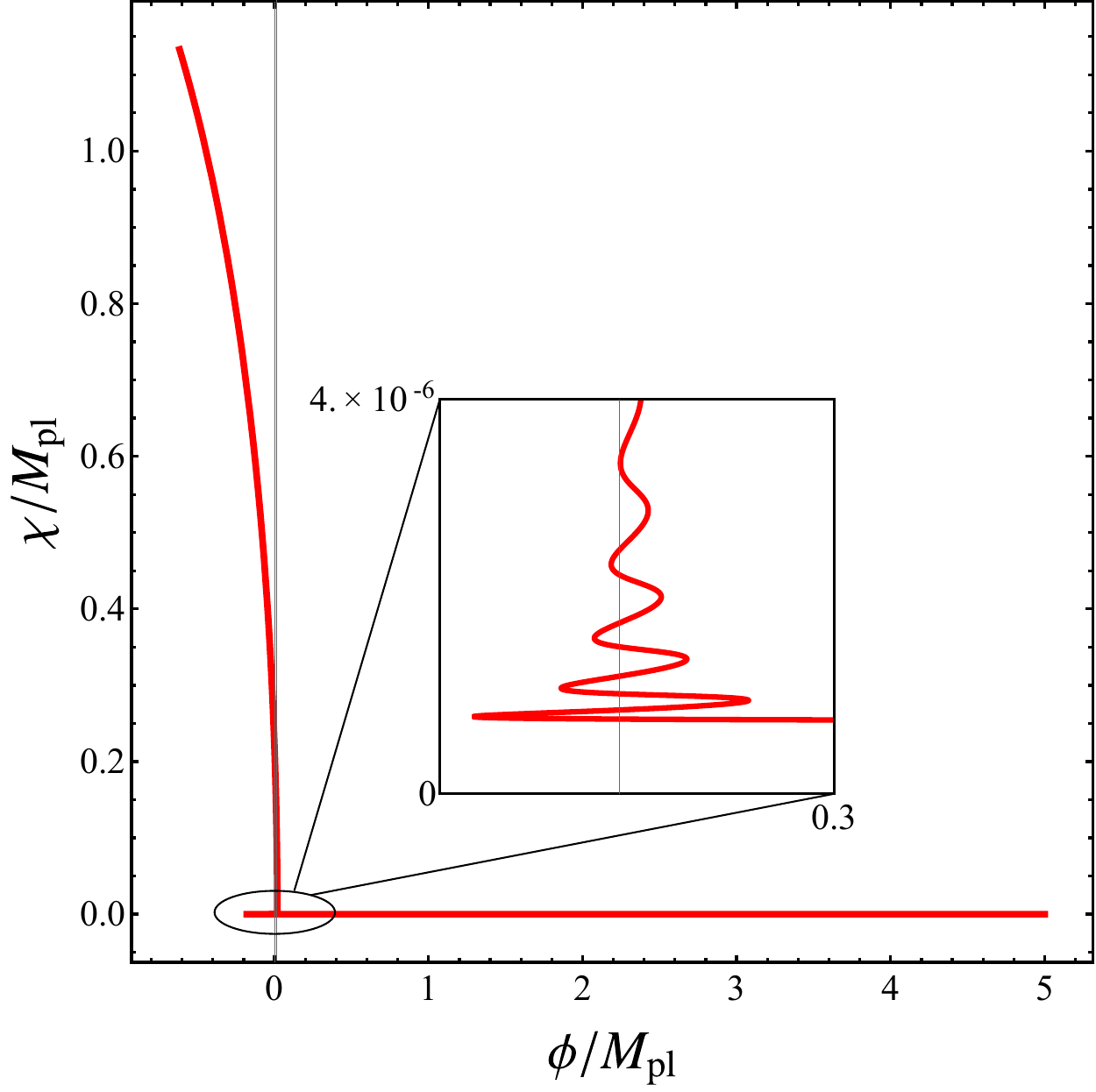}
\caption{The shape of the potential and the solution of the background equations (red lines). 
The inflaton first rolls down from the plateau along the $\phi$ direction at St-1, and begins its second slow-rolling along the $\chi$ direction at St-3. 
}
\label{potentialtrue}
\end{figure}

\begin{figure}[htbp]
\centering
\includegraphics[width=0.45\textwidth]{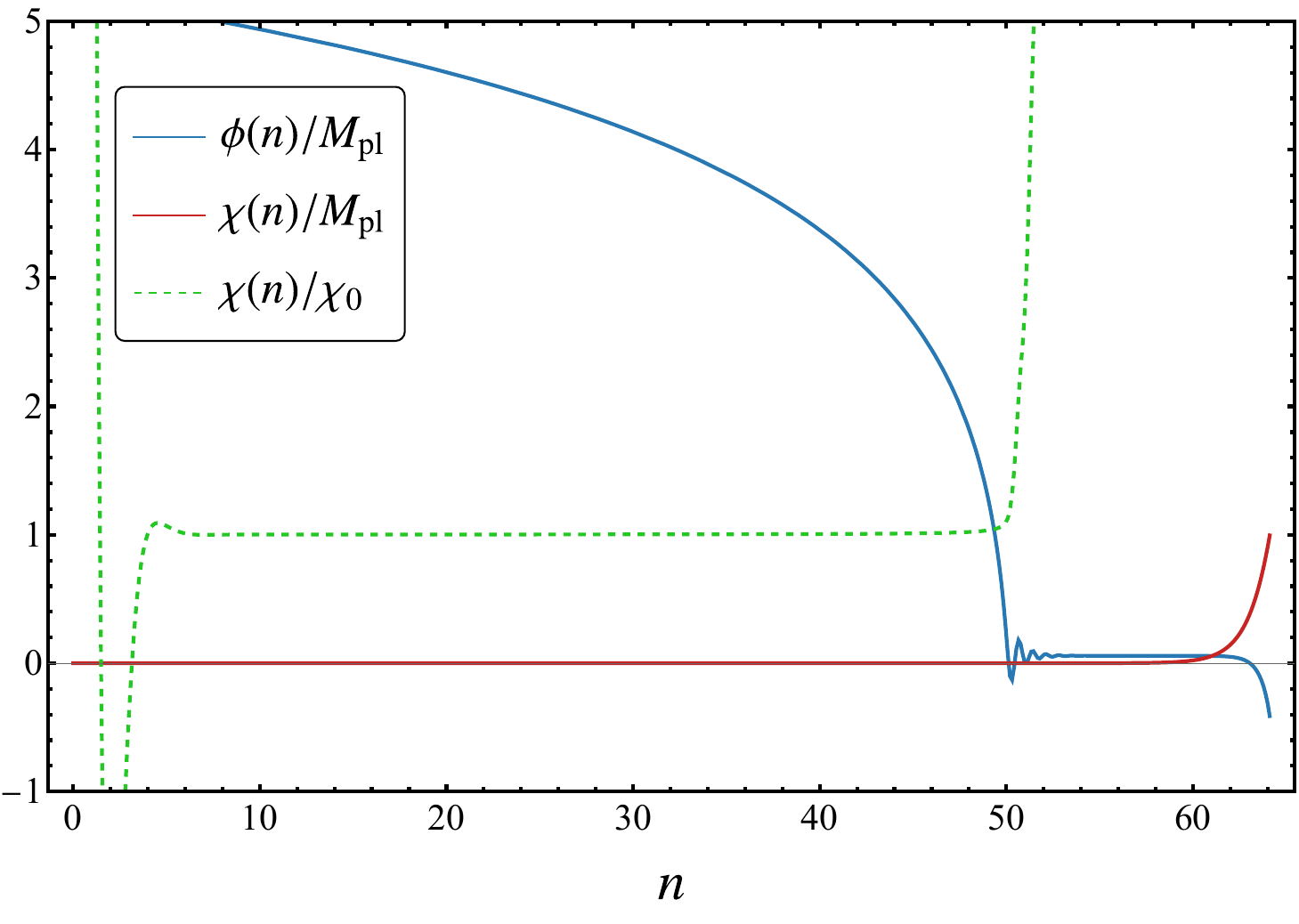}
\hspace{0.4in}
\includegraphics[width=0.45\textwidth]{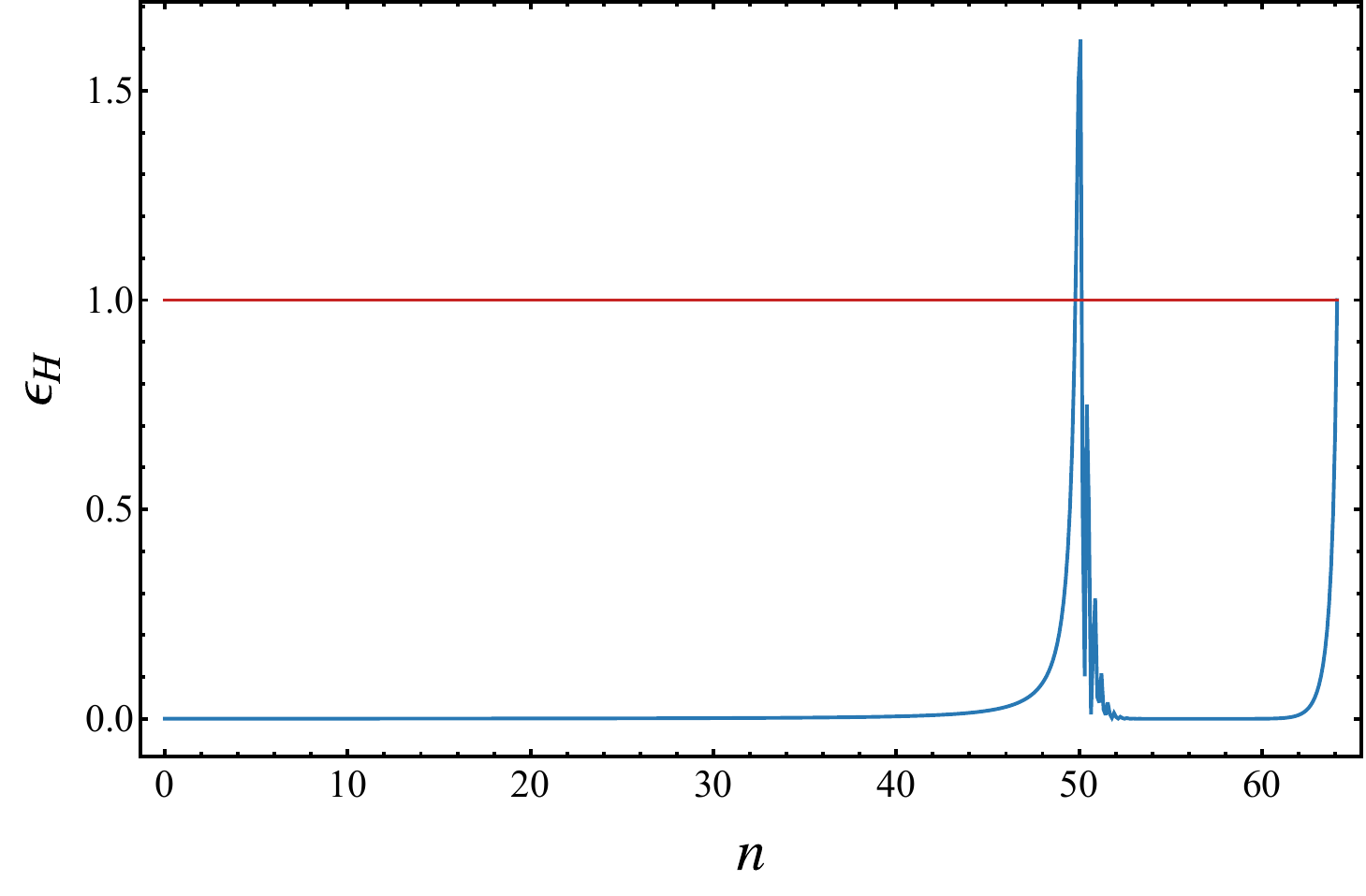}
\caption{
The background solution. The left panel shows the numerical solution for the background equations. The right panel shows the evolution of the slow-roll parameter $\epsilon_{H}$. It is obvious that during stage 2, the slow-roll condition is violated, hence resulting in a break between the two stages of inflation.}
\label{backgroundsolu}
\end{figure}

In the following, we work in the Einstein frame. Taking variation of Eq.\eqref{EinsteinS} with respect to $\tilde g^{\mu\nu}$, $\phi$ and $\chi$, assuming the Friedmann-Lemaitre-Robertson-Walker metric $\tilde g_{\mu\nu}=(-1, a^2(t), a^2(t), a^2(t))$, we obtain the Friedmann equations as well as the equations of motion for $\phi$ and $\chi$ as follows.
\begin{align}
&3M_\mathrm{pl}^2H^2=\frac{1}{2}\dot\phi^2+F^{-1}\frac{1}{2}\dot\chi^2+U(\phi,\chi)\,,\label{Fried1}\\
&-2M_{\mathrm{pl}}^2\dot H=\dot\phi^2+F^{-1}\dot\chi^2\,,\\
&
\begin{aligned}
&\ddot\phi+3H\dot\phi\\
& +\sqrt\frac{3}{2}M^2M_\mathrm{pl}F^{-1}\left\{FW(\phi,\chi)\left[1-W(\phi,\chi)\right]+\frac{1}{3}\left(MM_\mathrm{pl}\right)^{-2}\left[\dot\chi^2-4F^{-1}V(\chi)\right]\right\}=0\,,\label{phi1}
 \end{aligned}
 \\
    &\ddot{\chi}+\left(3H-\sqrt{\frac{2}{3}}\frac {\dot\phi}{M_\mathrm{pl}}\right)\dot\chi+F^{-1}\partial_{\chi}V+3\xi M^2(\chi-\chi_0)W(\phi,\chi)
=0\,,\label{chi1}
\end{align}
where a dot denotes the cosmic proper time derivative, $\dot~ \equiv\mathrm{d}/\mathrm{d}t$, 
and $H\equiv\dot a/a$ is the Hubble parameter. The inflationary universe in this model can be divided into three stages (see the potential feature and background solution in Fig.\ref{potential}, Fig.\ref{potentialtrue} and Fig.\ref{backgroundsolu}): 

\begin{itemize}[leftmargin=9mm]
\item[St-1.] At the first stage, as the effective mass of $\chi$ is much heavier than $H$, it quickly settles down to $\chi_0$. Since the value of $\phi$ that corresponds to the horizon exit of the CMB scale must be much larger than $M_{\rm pl}$ to maintain a sufficiently long slow-roll stage, we have $F(\phi)=\exp(\sqrt{2/3}\,\phi/M_{\mathrm{pl}})\gg1$. 
Therefore the Hubble parameter is dominated by the constant term $H\approx H_1= M/2$. 
During this stage, an effective single field inflation is realized, driven by the light scalaron $\phi$ on the flat plateau of $U(\phi, \chi)$. 
\item[St-2.] After the scalaron $\phi$ rolls down from the potential plateau, it undergoes damped oscillations around its local minimum at $\phi_c$, which is a function of $\chi$, and the first stage of inflation ends. In our scenario, as we consider the energy scale of the second inflationary stage to be much smaller than that of the first stage, it follows that $\phi_c\ll M_{\rm pl}$. Then due to the rapid decay of $\phi$ to $\phi_c$, $F(\phi)$ rapidly approaches unity, and the Hubble parameter $H$ drops down by several orders of magnitude.
As the rapid change of $H$ means the slow-roll parameter $\epsilon\equiv-\dot H/H^2$ of order unity, inflation is temporarily halted until the potential of $\chi$ starts to dominate. 
During this stage, the effective mass square of $\chi$ changes from positive to negative, which makes $\chi$ start rolling and drives the second stage of inflation. 
\item[St-3.] After $\phi$ settles down to $\phi_c$, the second inflationary stage begins, which is effectively single-field inflation dominated by $\chi$. During this stage, $\phi$ adiabatically evolves along $\phi=\phi_c(\chi)$, while $\chi$ rolls down from $\chi_0$ towards the minimum of the effective potential $V_{\rm eff}(\chi)=U(\phi_c(\chi),\chi)$, which is approximately the same as the minimum of $V(\chi)$, $\chi\approx\chi_g\equiv m/\sqrt{\lambda}$.
Finally inflation terminates at the global minimum of the potential, $(\chi,\phi)=(\chi_g,\phi_g)$ where $\phi_g=\sqrt\frac{3}{2}M_{\mathrm{pl}}\ln\left(1-\frac{\xi(m-\sqrt{\lambda}\chi_0)^2}{\lambda M_{\mathrm{pl}^2}}\right)$.
We note that, because of the potential shape and non-minimal coupling, this stage is similar to the Higgs inflation.
We also mention that the case $\lambda=\chi_0=0$ reduces to that discussed in \cite{Pi:2017gih} where neither the termination of inflation nor the issue of too large quantum fluctuations was taken into account.
\end{itemize}

Let us now comment on the role of $\chi_0$ in \eqref{fdef}. As explained above, or as shown in Fig.~\ref{potential}, $\chi$ is trapped at the local minimum $\chi\approx\chi_0$ during St-1. If $\chi_0=0$, as soon as St-2 starts at which $\phi$ settles down to the local minimum $\phi=\phi_c$, the point $\chi=0$ becomes a local maximum, implying that classically $\chi$ would be stuck $\chi=0$, and it would move away from $\chi=0$ only by the quantum fluctuations. This leads to unacceptably large curvature perturbations. A non-vanishing $\chi_0$ avoids this catastrophe.
The same strategy is also taken in \cite{Braglia:2022phb} to avoid the overproduction of topological defects which often happens in hybrid inflation scenarios such as \cite{Garcia-Bellido:1996mdl, Clesse:2015wea}. 
Before closing this section, let us briefly comment on the stability of the trajectory during St-1. During most of this stage, we have $F\gg1$ so that the effective mass of $\chi$ is positive and $\chi$ acts as a damped harmonic oscillator (See Fig.~\ref{potential}). In our model, we find that the coupling parameter $\xi$ must satisfy $\xi>3/16$ to ensure that $\chi$ undergoes an under-damped oscillator during St-1 so that it settles down to the stable local minimum $\chi_0$ in a few e-folds. This secures that St-1 can be effectively treated as single-field slow-roll inflation.\\

Hereafter, throughout the paper, we use the subscripts $\mathrm{i}$, $\star1$, $\star2$, and $\mathrm{f}$, respectively, to denote the quantities at the initial time our computations,
the initial time of the St-2 (or the end of St-1), the initial time of St-3 (or the end of St-2), and the end of inflation.

\section{The Background Solution}
\label{background}
We study the background solution in this section. It is convenient to introduce the e-folding number $n$ as the independent variable instead of the cosmological time $t$,
\begin{align}\label{Ndef}
    \mathrm{d}n=H\mathrm{d}t,
\end{align}
and denote the $n$-derivative by a prime, $'\equiv\mathrm{d}/\mathrm{d}n$. 
Then Eqs.\eqref{Fried1}--\eqref{chi1} are expressed in the form,
\begin{align}
    &H^2=\frac{U}{3-\epsilon_H},\qquad
    \epsilon_H\equiv-\frac{\dot H}{H^2}=\frac{1}{2M_{\mathrm{pl}}^2}\left(\phi'^2+\frac{1}{F(\phi)}\chi'^2\right),\label{H2N}\\ 
    &\phi''+\left(3+\frac{H'}{H}\right)\phi'\notag\\
    &+\sqrt\frac{3}{2}\frac{M^2M_\mathrm{pl}}{H^2}\left\{W(\phi,\chi)[1-W(\phi,\chi)]-\frac{4}{3M^2M_\mathrm{pl}^2F(\phi)^2}V(\chi)+\frac{H^2\chi'^2}{3M_{pl}^2M^2F(\phi)}\right\}=0,\label{phiN}\\
    &\chi''+\left(3-\frac{(F/H)'}{F/H}\right)\chi'+\frac{1}{H^2}\left(\frac{V_{,\chi}}{F(\phi)}+3\xi M^2(\chi-\chi_0)W(\phi,\chi)\right)=0,\label{chiN}
\end{align}
where $F$ and $W$ are given by Eqs.\eqref{phidef} and \eqref{Wdef}, respectively. 
Although we solve the above equations numerically in order to obtain sufficiently accurate results, it is useful to have an approximate analytical solution to better understand the behavior of the solution.
In the following, we solve for such a solution step by step from St-1 to St-3.

\subsection{St-1}
\label{stage1}
During St-1, the inflaton is slow-rolling along the $\phi$-direction. Since $\phi\gg M_{\rm pl}$, we have $F\gg 1$ and hence $H^2\approx H_1^2\equiv M^2/4$. Similarly, At St-3, since $\phi\simeq0$, we have $H^2\approx H_2^2\equiv V_0/(3M_{\mathrm{pl}}^2)$. 
We denote the ratio of $H_1$ to $H_2$ by $\mu$. Namely, 
\begin{align}\label{mudef}
    \mu^2\equiv\frac{H^2_1}{H^2_2}=\frac{3M^2M_{\mathrm{pl}}^2}{4V_0}=3\lambda\left(\frac{MM_{\mathrm{pl}}}{m^2}\right)^2,
\end{align} 
where we have used \eqref{V0mL} in the last equality. 
Then under the slow-roll approximation, \eqref{phiN} is approximated as
\begin{align}
     \frac{1}{M_\mathrm{pl}}\frac{\mathrm{d}\phi}{\mathrm{d}n}&\approx2\sqrt\frac{2}{3}\frac{\left(1+\mu^{-2}\right)-F(\phi)}{(F(\phi)-1)^2+\mu^{-2}}.
     \label{phieq1}
\end{align}
If we further assume $F\gg1$ ($\phi/M_{\rm pl}\gg1$) and $\mu^2\gg1$, the above reduces to
\begin{align}
 \frac{1}{M_{\rm pl}}\frac{d\phi}{dn}\approx-2\sqrt{\frac{2}{3}}\frac{1}{F(\phi)}
 \quad\leftrightarrow\quad\frac{dF}{dn}\approx -\frac{4}{3}\,.
\label{dphisol1}
\end{align}
Thus $F$ rapidly decreases in comparison with the slow motion of $\phi$.

The first stage of slow-roll inflation ends at the point when $\epsilon_H\simeq\left.\phi'^2/2\right|_{\star}=1$. 
Inserting this condition into \eqref{phieq1}, we obtain the value of $\phi$ at the end of St-1,
\begin{align}\label{phistar}
\phi_{\star1}\equiv\phi(n_{\star1})
=\sqrt{\frac{3}{2}}M_\mathrm{pl}\ln\left(1+\frac{\sqrt{3}}{3}+\frac{\sqrt{3}}{3}\sqrt{1-\frac{3+2\sqrt{3}}{\mu^2}}\right)\,.
\end{align}
Apparently, we must have $\mu^2>3+2\sqrt{3}\simeq6.46$ in order for our scenario to be sensible. 
As we are interested in the case in which inflation temporarily halts, we consider the limit $\mu^2\gg 1$.
So this constraint on $\mu^2$ does not play any significant role. 
In this limit, for $\phi/M_{\rm pl}\gg 1$, we obtain \eqref{dphisol1}. 
We also mention that in the limit $\mu^2\gg1$, we have $\phi_{\star1}\approx0.94 M_{\rm pl}$.
Then \eqref{dphisol1} can be easily solved to give
\begin{align}\label{phiN1}    \phi(n)=M_{\mathrm{pl}}\sqrt{\frac{3}{2}}\ln\left[\frac{4}{3}(n_{\star1}-n)+F_{\star1}\right]
\quad
\leftrightarrow\quad
F(n)=F_{\star 1}+\frac{4}{3}(n_{\star1}-n)\,,
\end{align}
where $F_{\star1}$ is the value of suppression factor at the end of St-1, $n=n_{\star1}$,
\begin{align}\label{Fstar1}
F_{\star1}\equiv F(\phi(n_{\star1}))=1+\frac{1}{\sqrt{3}}\left(1+\sqrt{1-\frac{3+2\sqrt{3}}{\mu^2}}\right)\approx
1+\frac{2}{\sqrt{3}}.
\end{align}
The numerical calculations of $\phi$ and $\epsilon_H$ are shown in Fig.~\ref{backgroundsolu}.  
Taking the limit $\mu^2\gg1$ in \eqref{phieq1}, but without assuming $F\gg1$, we have
\begin{align}
     {M_\mathrm{pl}}\frac{\mathrm{d}n}{\mathrm{d}\phi}&\approx-\frac{1}{2}\sqrt\frac{3}{2}\left(F(\phi)-1\right)\quad \rm{for}\ \mu^2\gg 1.
     \label{nstage1evo}
\end{align}
By integrating this equation, the duration of St-1 from the initial condition $\phi=\phi_\mathrm{i}$ is estimated as 
\begin{align}
    N_{1}(\phi_\mathrm{i})=n_{\star}-n_\mathrm{i}&=\frac{3}{4}\left[(F(\phi_{\mathrm{i}})-\ln F(\phi_{\mathrm{i}}))-(F_{\star1}-\ln F_{\star1})\right]
    \nonumber\\
    &\approx\frac{3}{4}(F(\phi_{\mathrm{i}})-F_{\star1}+\ln F_{\star 1})\,,
   \label{N1}
\end{align}
where the second approximate equality follows when $F(\phi_i)\gg1$.

Now we turn to the motion of $\chi$. As we mentioned in Section \ref{model}, we assume the non-minimal coupling constant satisfied $\xi>3/16$ so that $\chi$ 
has a large effective mass, $m_{\chi,\rm eff}^2\approx3\xi M^2>9H^2/4\approx 9M^2/16$. Thus during St-1, it is trapped in the valley of $U$ along the $\phi$-direction, as illustrated in the upper-left panel of Fig.~\ref{potential}. Assuming $F\gg1$, the value of $\chi$ at the bottom of the valley can be easily evaluated from \eqref{chi1} or \eqref{chiN}. 
\begin{align}
  \left. \frac{\partial U}{\partial\chi}\right|_{\chi=\chi_{c}}
  =&{3M^2\xi(\chi_{c}-\chi_0)}F^{-1}\notag\\
  &+\left(3M^2M_\mathrm{pl}^{-2}\xi^2(\chi_{c}-\chi_0)^3-3M^2\xi(\chi_{c}-\chi_0)+\lambda\chi_{c}^3-m^2\chi_{c}\right)F^{-2}=0\,.
  \label{chic}
\end{align}
We find 
\begin{align}\label{chicapp}
  \chi_{c}=\chi_0+O(F^{-1})\,,
\end{align}
that is, we may approximate $\chi_c$ by $\chi_0$.
Since $\chi$ quickly approaches $\chi_c$ regardless of its initial value, we may treat $\Delta\chi\equiv\chi-\chi_c$ as a small deviation from $\chi_c$. 
Up to linear order, the equation of motion for $\Delta\chi$ can be approximated as
\begin{align}\label{deltachi}
\ddot{\Delta\chi}+\left(3H-\frac{\dot F}{F}\right)\dot{\Delta\chi}+m_{\chi,\mathrm{eff}}^2\Delta\chi=0\,,
\end{align}
where the exact expression for $m_{\chi,\mathrm{eff}}^2$ is given by
\begin{align}\label{meff}
     m_{\chi,\mathrm{eff}}^2=F\frac{\partial^2U}{\partial\chi^2}=3M^2\xi+F^{-1}\left[9M^2\xi(\chi-\chi_0)^2M_{\mathrm{pl}}^{-2}+3\lambda\chi^2-3M^2\xi-m^2\right].
\end{align}
The second term proportional to $F^{-1}$ is negligible during most of St-1. However, its contribution starts to dominate when St-1 ends as $F$ becomes of order unity there. 
Note that we took into account the time variation of $F/H$, which is small but may have an appreciable secular effect.
Adopting the approximation $m_{\chi,\mathrm{eff}}^2\approx3\xi M^2$, \eqref{deltachi} can be solved to give
\begin{equation} \label{chisol1}
\chi(n)-\chi_0=e^{-\frac{3}{2}n}\left(\frac{F/H}{F_\mathrm{i}/H_\mathrm{i}}\right)^{1/2}
\left[(\chi_{\mathrm{i}}-\chi_0)\cos\left(\omega n\right)
+\dfrac{\chi_i'}{\omega}\sin\left(\omega n\right)\right]\,; ~\omega\equiv\frac{3}{2}\sqrt{\frac{16}{3}\xi-1}\,,
 \end{equation}
 where we have set $n_{i}=0$ for simplicity. As clear from this expression, $\chi(n)=\chi_0$ in high accuracy for $n\gg1$. However, for $n\lesssim 1$, the difference between $\chi(n)$ and $\chi_0$ becomes appreciable. 
Although this won't change the number of e-folds of St-1 given by \eqref{N1}, it will make a difference in the number of e-folds of St-3 as we shall see below.

\subsection{St-2}
St-2 is the intermediate stage where inflation is halted for a while and the inflaton turns the direction of its trajectory from $\phi$ to $\chi$-direction. During this stage, $\chi$ becomes destabilized and starts to roll down the potential hill slowly, while
$\phi$ begins damped oscillations around its local minimum $\phi_c$. We note that the energy density of the oscillations is responsible for the break of inflation. Let us first consider the ranges of the parameters of the model that successfully realize our scenario. 

Using the relation $\left(\partial U/\partial\phi\right)|_{\phi=\phi_c, \chi=\chi_0}=0$, together with $M^2M_{\mathrm{pl}}^2\gg V_0$ from \eqref{mudef}, we obtain 
\begin{align}\label{phic}
    \frac{\phi_c}{M_{\mathrm{pl}}}\approx\sqrt{\frac{3}{2}}\frac{1}{\mu^2}\ll1,
\end{align}
where $\mu^2$ is defined in \eqref{mudef}. Here we note that, in order to realize the second inflationary stage, St-3, $\chi_0$ must satisfy the constraint,
\begin{align}\label{chi0upper}
    \chi_0^2<\frac{m^2}{\lambda}=\chi_g^2.
\end{align}
At the beginning of St-2, we make sure $\chi$ is still trapped in the valley. Hence its effective mass square should be positive.  
From \eqref{meff}, we have
\begin{align}
     \left.m_{\mathrm{eff},\chi}^2\right|_{\star1}
     &=3M^2\xi+F_{\star1}^{-1}\left[3A\xi m^2(\chi_0/M_{\rm{pl}})^2-3M^2\xi-m^2\right]
     \nonumber\\
     &=3M^2\xi\left[1-\frac{1+A\mu^{-2}}{1+\frac{1}{\sqrt{3}}\left(1+\sqrt{1-
     ({3+2\sqrt{3}})}\mu^{-2}\right)}+\mathcal{O}\left(\frac{m^2}{M_{\rm pl}^2}\right)\right]>0\,,
\end{align}
where we have introduced, the parameter $A$ defined by
\begin{eqnarray}\label{Adef}
   A\equiv\lambda \frac{M_{\mathrm{pl}}^2}{\xi m^2}\,.
\end{eqnarray}
This requires $A<\mu^2\left(1+\sqrt{1-(3+2\sqrt{3})/\mu^2}\right)/\sqrt{3}$.
On the other hand, at the end of St-2, the effective mass square of $\chi$ should be negative so that it can start rolling down and realize inflation.
Thus we require
\begin{align}
     \left.m_{\mathrm{eff},\chi}^2\right|_{\star2}&
     =3M^2\xi+F(\phi_c)^{-1}\left[3A\xi m^2(\chi_0/M_{\rm{pl}})^2-3M^2\xi-m^2\right]
     \nonumber\\
     &= 3M^2\xi\left[1-\frac{1+A\mu^{-2}}{1+\mu^{-2}}+\mathcal{O}\left(\frac{m^2}{M_{\rm pl}^2}\right)\right]<0\,,
\end{align}
hence $A>1$, where we have used the fact that $\phi_{\star2}=\phi_c$ and $\chi_{\star2}=\chi_0$. 
To summarize, to realize our scenario of inflation, we require the parameter $A$ to satisfy
\begin{align}\label{lambdaupper}
   1<A<\frac{\mu^2}{\sqrt{3}}\left(1+\sqrt{1-(3+2\sqrt{3})/\mu^2}\right)\approx\frac{2}{\sqrt{3}}\mu^2\,.
\end{align}
We note that, in terms of $A$, we may re-express $\mu^2$ and $\chi_g^2$ as $\mu^2=3A\xi(M^2/m^2)$ and $\chi_g^2=M_{\mathrm{pl}}^2(A\xi)^{-1}$.

%

 \begin{figure}[htbp]
 \centering
\includegraphics[width=0.7\textwidth]{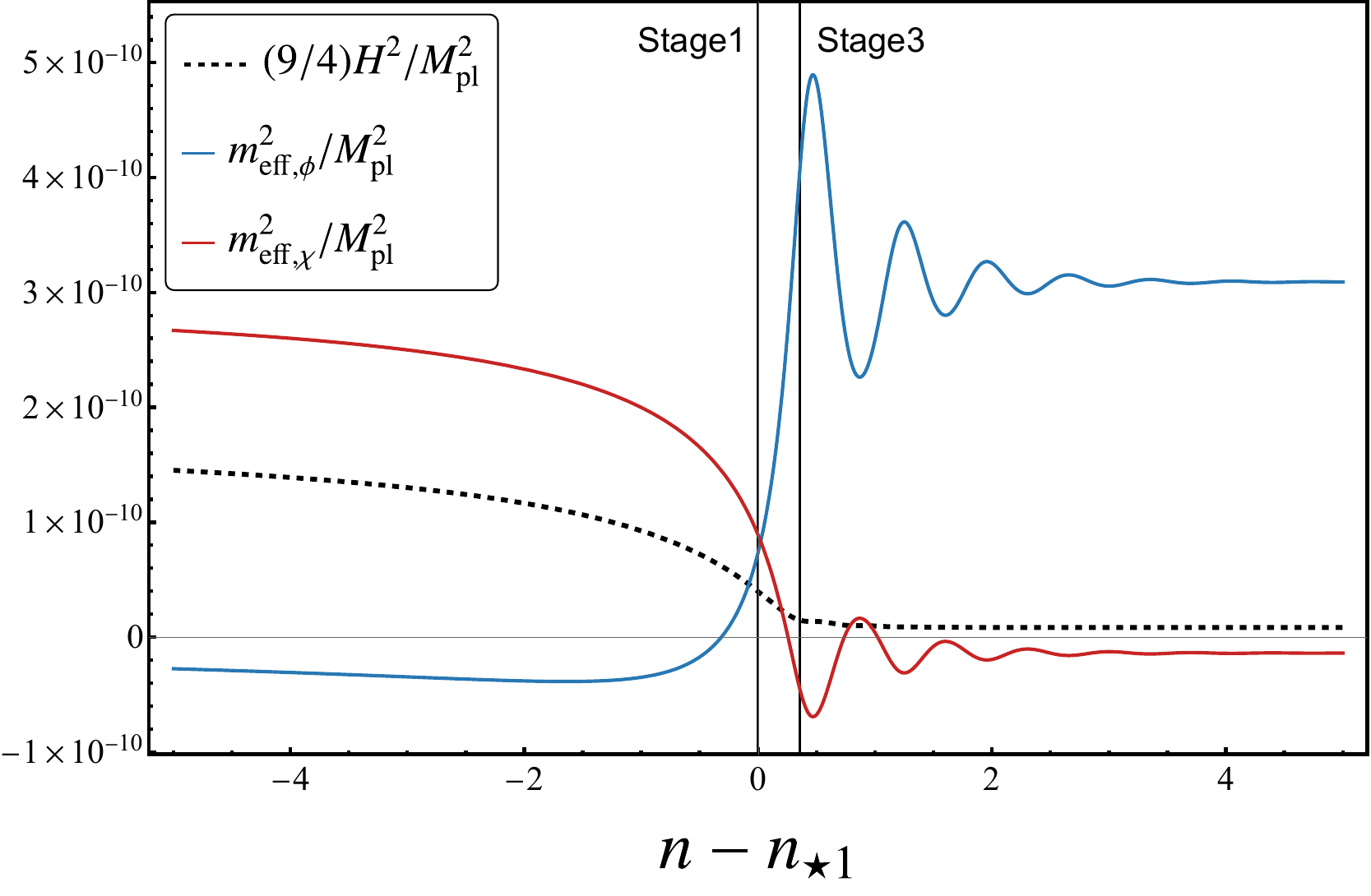}
\caption{The numerical result of the effective mass of the two scalar fields. The blue, red solid line and the black dotted line show the time evolution of the effective mass of $\phi$, the effective mass of $\chi$, and $H^2$. The parameters used for this plot can be found in case 1 of Table. \ref{parameters}. The two vertical solid lines indicate from the left to right the horizon exiting time for modes $k_2$ and $k_1$.}
\label{massplot}
\end{figure}
Now we turn to solving the evolution. 
We show the plot of the time evolution of the effective mass square for $\chi$ and $\phi$ with a typical choice of the model parameters in Fig.~\ref{massplot}. During St-1, $\phi$ is the light field that drives inflation, $m_{\phi,\rm eff}^2<(9/4)H^2$, while $\chi$ is heavy and stably stays around $\chi_0$, with the deviation $\chi-\chi_0$ decays rapidly as $a^{-3/2}$.  
During St-2, the roles of $\chi$ and $\phi$ are interchanged. The field $\chi$ is destabilized from $\chi_0$ and starts rolling. As for $\phi$, it becomes heavy and undergoes damped oscillations around $\phi_c$. Similar to the case of $\chi$ at St-1 given by \eqref{deltachi}, setting $\Delta\phi\equiv\phi-\phi_c$,  the equation of motion \eqref{phi1} can be approximated as
\begin{align}\label{deltaphieom}    
\ddot{\Delta\phi}+3H\dot{\Delta\phi}+M^2\Delta\phi=0\,.
\end{align}
To make sure $\phi$ is indeed heavy, it may be useful to estimate the Hubble parameter at the beginning of St-2.
Since $H$ always decreases in time, we obtain
\begin{align}\label{hstar}
H^2<H_{\star1}^2&\approx\frac{U(\phi_{\star1}, \chi_{\star1})}{2M_{\mathrm{pl}}^2}\approx\frac{3}{8}M^2\left(\left(1-F_{\star1}^{-1}\right)^2+\mu^{-2}F_{\star1}^{-2}\right)
\nonumber\\
&\lesssim\frac{1}{8}\left(\frac{2}{1+2\cdot3^{-1/2}}\right)^2M^2\simeq0.1M^2\,,
\end{align}
where we have neglected $V(\chi)$ in $U(\phi,\chi)$ as it is still negligible in the beginning of St-2. 
Thus we obtain the solution,
\begin{align}
    \Delta\phi(t)\approx(\phi_{\star1}-\phi_c)\left(\frac{a(t_{\star1})}{a(t)}\right)^{-{3}/{2}}\cos\left(M(t-t_{\star1})\right),
    \label{soluphi2}
\end{align}
where we have ignored the initial velocity at $t=t_{\star1}$ for simplicity.
During this stage, the potential $U$ is evaluated as 
\begin{align}\label{approxH1}
    U\approx\frac{1}{2}M^2\phi^2+V_0\approx\frac{1}{2}M^2\Delta\phi^2+V_0\,,
\end{align}
where in the last step we used \eqref{phic}, i.e., the fact that $\phi_c/M_{\rm pl}\ll1$, hence $\phi=\Delta\phi+\phi_c\approx\Delta\phi$. 
As for $\chi$, as it has only started to roll slowly, its kinetic energy is negligible. 
Therefore, the Hubble parameter, which is the sum of the kinetic and potential energies, is given by 
\begin{align}
    H^2\approx\frac{1}{3M_{\mathrm{pl}}^2}\left[\frac12\left(\dot{\Delta\phi}^2+M^2\Delta\phi^2\right)+V_0\right]
    \approx\frac{1}{3M_{\mathrm{pl}}^2}\left[\frac{M^2\phi_{\star1}^2}{2}\frac{a^3(t_{\star1})}{a^3(t)}+V_0\right],
    \label{Hamp}
\end{align}
where in the last step we have used $\phi_{\star1}-\phi_c\approx\phi_{\star1}$.

From the definition of the slow-roll parameter $\epsilon_H$ given in \eqref{H2N}, it is estimated as
\begin{align}
\epsilon_H=3-\frac{U}{M_{\mathrm{pl}}^2H^2}\approx
\frac{3(a_{\star1}/a)^{3}M^2\phi_{\star1}^2\sin^2(M(t-t_{\star1}))}{(a_{\star1}/a)^{3}M^2\phi_{\star1}^2+2V_0}\,.
\end{align}
Since $H^2\ll M^2$, $\epsilon_H$ oscillates many times if the duration of St-2 is one e-fold or more.
Note that the time-averaged amplitude of $\epsilon_H$, denoted by $\langle\epsilon_H\rangle$, is approximately $3/2$ in the beginning.  
St-2 terminates when it decreases to below unity. As there is no exact definition of the end of St-2, we set this critical value of
$\langle\epsilon_H\rangle$ to $1/p$ where $p>1$. Then
\begin{align}
\left\langle\epsilon_H\right\rangle_{\star2}
=\frac{3(a_{\star1}/a)^{3}M^2\phi_{\star1}^2}{2(a_{\star1}/a)^{3}M^2\phi_{\star1}^2+4V_0}\bigg|_{\star2}=\frac{1}{p}
\quad\Rightarrow\quad \frac{a_{\star2}}{a_{\star1}}=\mu^{\frac23}\left(\frac{(3p-2)\phi_{\star1}^2}{3M_{\mathrm{pl}}^2}\right)^{\frac{1}{3}},
\label{asolu}
\end{align}
 we have used \eqref{mudef} to express $V_0$ and $M^2$ in terms of $\mu^2$. 
Thus the total e-folding number of St-2 is estimated as
\begin{align}
N_2=\frac{1}{3}\ln\left[\frac{3p-2}{3}\mu^2\left(\frac{\phi_{\star1}}{M_{\mathrm{pl}}}\right)^2\right]\simeq\frac{1}{3}\ln{\mu^2},
\label{N2}
\end{align}
for $1\lesssim p\lesssim 3$ and $\mu^2\gg1$, where we have used the fact that $\phi_{\star1}\simeq 0.94M_{\rm pl}$ from \eqref{phistar}. 
We obtain $N_2\gtrsim1$ for $\mu^2\gtrsim20$, which justifies our assumption that St-2 is long enough to accommodate many oscillations of $\phi$. 

The amplitude of $\phi$ at the end of St-2 may be evaluated by using \eqref{soluphi2} and \eqref{asolu},
\begin{align}
    \phi_{\star2}=\phi_c+(\phi_{\star1}-\phi_c)\left(\frac{a_{\star1}}{a_{\star2}}\right)^{\frac{3}{2}}
    =\phi_c+\frac{\sqrt{3/(2p-2)}}{\mu}M_{\mathrm{pl}}\left(1-\frac{\phi_c}{\phi_{\star1}}\right)\simeq\phi_c\,,
\end{align}
which we can use as the initial value of $\phi$ at the beginning of St-3.

Now we turn to solve the evolution of $\chi$. In the beginning, $\chi$ stays at the bottom of the potential valley sustained by the positive effective mass square, $m_{\chi,\rm eff}^2>0$, which starts to decrease as the amplitude of $\phi$ approaches $\phi_c$. However, $m_{\chi,\rm eff}^2$ becomes negative only near the end of St-2, and since $|m_{\chi,\rm eff}^2|\ll H^2$ in order to realize slow-roll inflation in the direction of $\chi$, it virtually does not start rolling at all during St-2. Therefore, as a leading order approximation, we may assume that $\chi$ stays at $\chi_0$ during St-2.
Thus the initial condition for St-3 is given simply by
\begin{align}
    \chi_{\star2}=\chi(n_{\star2})\simeq\chi_0\,.
    \label{t23}
\end{align}
We note that, however, it is important to consider the neighboring trajectories off from the fiducial one for the computations of the curvature perturbation based on the $\delta N$ formalism. 
Thus $\chi_{\star 2}$ is close to $\chi_0$ but the difference matters.

\subsection{St-3}
During St-3, the inflaton starts slow-rolling along the $\chi$-direction, while continuing to oscillate along the $\phi$-direction around its local minimum $\phi_c(n)$ which slowly evolves with time, given by $\left(\partial{U}/\partial\phi\right)|_{\phi=\phi_c(n)}=0$, from which we obtain
\begin{align}
\label{phicN}
    F(\phi_c)=&\exp\left[\sqrt{\frac{2}{3}}\frac{\phi_c}{M_{\rm pl}}\right]=
    1-\xi\left(\frac{\chi-\chi_0}{M_{\mathrm{pl}}}\right)^2
    \nonumber\\
   & +\mu^{-2}\left[1-\xi\left(\frac{\chi-\chi_0}{M_{\mathrm{pl}}}\right)^2\right]^{-1}\left[1-\left(\frac{\chi}{\chi_g}\right)^2\right]^2,
\end{align}
where $\chi_g\equiv m/\sqrt{\lambda}$ is the value of $\chi$ at the global minimum, firstly mentioned below \eqref{Vchi}. 
It is evident that $\phi_c$ evolves along with the evolution of $\chi$, as illustrated in the lower right panel of Fig.~\ref{potential}. 
Similar to the way we derived \eqref{soluphi2}, we set $\phi(n)=\phi_c(n)+\Delta\phi(n)$ with $\Delta\phi(n)\ll\phi_c(n)$ during St-3. 
Inserting this ansatz into \eqref{phi1}, neglecting the contribution of $\xi(\chi-\chi_0)^2/M_{\mathrm{pl}}^2$ and $\chi/\chi_g$ terms in the regime where $\chi\ll \chi_g\ll M_{\rm pl}$, the equation of motion for $\Delta\phi(n)$ can be approximated as
\begin{align}\label{deltaphiapp}
\ddot{\Delta\phi}+3H\dot{\Delta\phi}+\left(M^2+8H^2\right)\Delta\phi=0\,,
\end{align}
where $H^2\approx H_2^2=V_0/(3M_{\rm pl}^2)$.
The solution is
\begin{align}    \Delta\phi&=\Delta\phi_{\star2} \exp\left({-\frac{3}{2}H_2(t-t_{\star2})}\right)\cos\left[\sqrt{M^2+8H_2^2}(t-t_{\star2})\right]\nonumber\\
&\approx\Delta\phi_{\star2} \exp\left({-\frac{3}{2}(n-n_{\star2})}\right)\cos\left[2\mu(n-n_{\star2})\right],
\end{align}
where we have used $\mu^2\gg1$ in the last step. 

In order to obtain the approximate solution for $\chi$, we let us first evaluate the potential term in \eqref{chiN}, 
\begin{align}
    F(\phi_c)\frac{\partial U}{\partial\chi}\frac{1}{H^2}&=-\frac{36M^2M_{\mathrm{pl}}^2(1-(\chi/\chi_g)^2)(\chi M_{\mathrm{pl}}^2/\chi_g^2-\xi(\chi-\chi_0)((\chi/\chi_g)^2+1))}{3M^2(M_{\mathrm{pl}}^2-\xi(\chi-\chi_0)^2)^2+m^2\chi_g^2\left(\left({\chi}/{\chi_g}\right)^2-1\right)^2}\nonumber\\
    &\simeq-12\xi\left[(A-1)\chi+\chi_0\right],
    \label{dudchi}
\end{align}
where we have used the relation 
$A=\lambda M_{\mathrm{pl}}^2/(\xi m)^2=M_{\mathrm{pl}}^2/(\xi\chi_g^2)$
and neglected the $(\chi/\chi_g)^2$ and $\mu^{-2}$ terms.
The remaining term proportional to $W$ is negligible since $W\simeq (F-1)/F\ll1$.
Hence, we have
\begin{align}
    \chi''+3\chi'-12\xi\left[(A-1)\chi+\chi_0\right]=0 ,
\end{align}
by defining
\begin{align}  &\beta\equiv3\left(\sqrt{1+\frac{16}{3}\xi(A-1)}-1\right)\,.
\end{align}
The solution is readily obtained as
\begin{align}\label{chi3sol}
\chi(n)&=-\frac{\chi_0}{A-1}
 +\left(\chi_\mathrm{\star_2}+\frac{\chi_0}{A-1}\right)
 \exp\left[\frac{\beta}{2}(n-n_{\star2})\right]
 \nonumber\\
 &\approx
 -\frac{\chi_0}{A-1}
 +\left(\frac{A}{A-1}\right)\chi_0
 \exp\left[4\xi(A-1)(n-n_{\star2})\right],
\end{align}
where we have assumed the slow-roll condition $\chi''\ll\chi'$, which demands
\begin{align}\label{xiAapp}
\big|\xi(A-1)\big|\ll1\,.
\end{align}
Under this condition, slow-roll inflation is successfully realized during St-3. 

The end inflation is determined by $\epsilon_H=1$, 
\begin{align}\label{srH3}
   \epsilon_H\approx\frac{\chi'^2}{2M_{\mathrm{pl}}^2}\bigg|_{n=n_{\mathrm{f}}}=1.
\end{align}
Taking the $n$-derivative of \eqref{chi3sol}, the above determines the e-folding number $n_{\rm f}$ at the end of inflation. Then inserting \eqref{chi3sol} to \eqref{srH3} by setting $n=n_{\rm f}$, we obtain the duration of St-3 as
\begin{align}\label{ends3}
  N_3&=n_{\mathrm{f}}-n_{\star2}
  =\frac{2}{\beta}
  \ln\left[\frac{2\sqrt2 M_{\rm pl}}{\left(\chi_{\star2}+\frac{\chi_0}{A-1}\right)\beta}\right]
  \nonumber\\
  &\approx\frac{1}{4\xi(A-1)}\ln\left[\frac{\sqrt{2}M_{\rm pl}}{4\xi A\chi_0+(A-1)\Delta\chi_{\star 2}}\right]
  \nonumber \\
 &=\frac{1}{4\xi(A-1)}
 \left(\ln\left[\frac{\sqrt{2}}{4}\left(\frac{\chi_g}{M_{\mathrm{pl}}}\right)\left(\frac{\chi_g}{\chi_0}\right)\right]-\ln\left[1+\dfrac{(A-1)\chi_g^2}{4M_{\rm pl}^2}
\frac{\Delta\chi_{\star 2}}{\chi_0}\right]\right)\,,
\end{align}
where $\Delta \chi_{\star 2}=\chi_{\star 2}-\chi_0$.
As we must have $\chi_0<\chi(n_{\mathrm{f}})<\chi_g$, $\chi_0$ is limited by 
\begin{align}
\chi_g\left[1+A\left(\frac{\sqrt{2}}{4}\frac{\chi_g}{M_{\mathrm{pl}}}-1\right)\right]
<\chi_0<\frac{\sqrt{2}}{4}\frac{\chi_g}{M_{\rm pl}}\chi_g\,,
\end{align}
where we have ignored the contribution of $\Delta\chi_{\star 2}$ as $\Delta\chi_{\star 2}/\chi_0\ll 1$.
It turns out the left inequality is automatically satisfied unless $A<1$, provided that the right inequality is satisfied, i.e., $\sqrt{2}\chi_g/(4M_{\rm pl})<1$. 

Finally, let us check our assumption $\dot\chi^2\ll V_0$ which is used when deriving the equation of motion \eqref{deltaphiapp}. Using \eqref{chi3sol}, we obtain 
\begin{align}\label{chidotV}
\frac{\dot\chi^2}{V_0}=\frac{H_2^2\chi'^2}{V_0}&\approx\frac{16}{3}\left(\frac{\chi_0}{\chi_g}\right)^2\left(\frac{M_{\mathrm{pl}}}{\chi_g}\right)^2\exp\left[8\xi(A-1)(n-n_{\star2})\right]<\frac{2}{3},
\end{align}
where we used $n-n_{\star2}<n_{\mathrm{f}}-n_{\star2}$. This justifies the validity of our approximation.



\section{Perturbation}
\label{perturbation}
\subsection{Preliminaries}
In the case of single-field slow-roll inflation, the superhorizon behavior of the adiabatic growing mode is simple. The comoving curvature perturbation stays constant, $\mathcal{R}_c=\mathrm{const.}$
However, in multi-field inflation, $\mathcal{R}_c$ varies in time even if all the fields are slow-rolling and the modes are far outside the horizon. In the language of the separate universe approach, this is because there are many possible trajectories (histories) of the universe in the phase space, and quantum fluctuations can induce jumps from one trajectory to another. Given a fiducial trajectory, it is customary to attribute this time evolution to ``isocurvature modes $\mathcal S$", because they are the modes that are orthogonal to the fiducial trajectory. 
The comoving curvature and isocurvature perturbations are expressed as follows:
\begin{align}
    &\mathcal{R}_c(n)=\frac{\phi'(n)\delta\phi(n)+F(\phi(n))^{-1}\chi'(n)\delta\chi(n)}{2\epsilon_H(n)M_{\mathrm{pl}}^2},\\
    &\mathcal{S}(n)=F(\phi(n))^{-1/2}\frac{\chi'(n)\delta\phi(n)-\phi'(n)\delta\chi(n)}{2\epsilon_H(n)M_{\mathrm{pl}}^2},
    \label{cands}
\end{align}
where $\delta\phi$ and $\delta\chi$ are those defined on the flat slicing. The detailed derivation of these expressions is given in Appendix B.

Except for the transient stage, ie, toward the end of St-1 and during St-2, when the coupling between $\delta\phi$ and $\delta\chi$ may become important, their behavior on superhorizon scales may be separately discussed as both St-1 and St-3 may be regarded as effectively single-field slow-roll inflation. Thus, the curvature perturbation is dominated by $\delta\phi$ at St-1, while it is dominated by $\delta\chi$ at St-3.
The isocurvature perturbation does not play significant roles at both St-1 and St-3, as $\chi'$ ($\phi'$) is negligible and $\delta\chi$ ($\delta\phi$) decays rapidly at St-1 (St-3).  
The situation is quite different during the transient stage. Toward the end of St-1, $\delta\phi$ becomes massive and its contribution to the curvature perturbation starts to decrease. On the other hand, the isocurvature perturbation is enhanced and becomes the same order of magnitude as the curvature perturbation. This large amplitude isocurvature component transforms into the curvature perturbation as the background trajectory turns to the $\chi$-direction. This induces the evolution of the curvature perturbation on superhorizon scales.

\begin{figure}[htbp]
\centering
\includegraphics[width=0.7\textwidth]{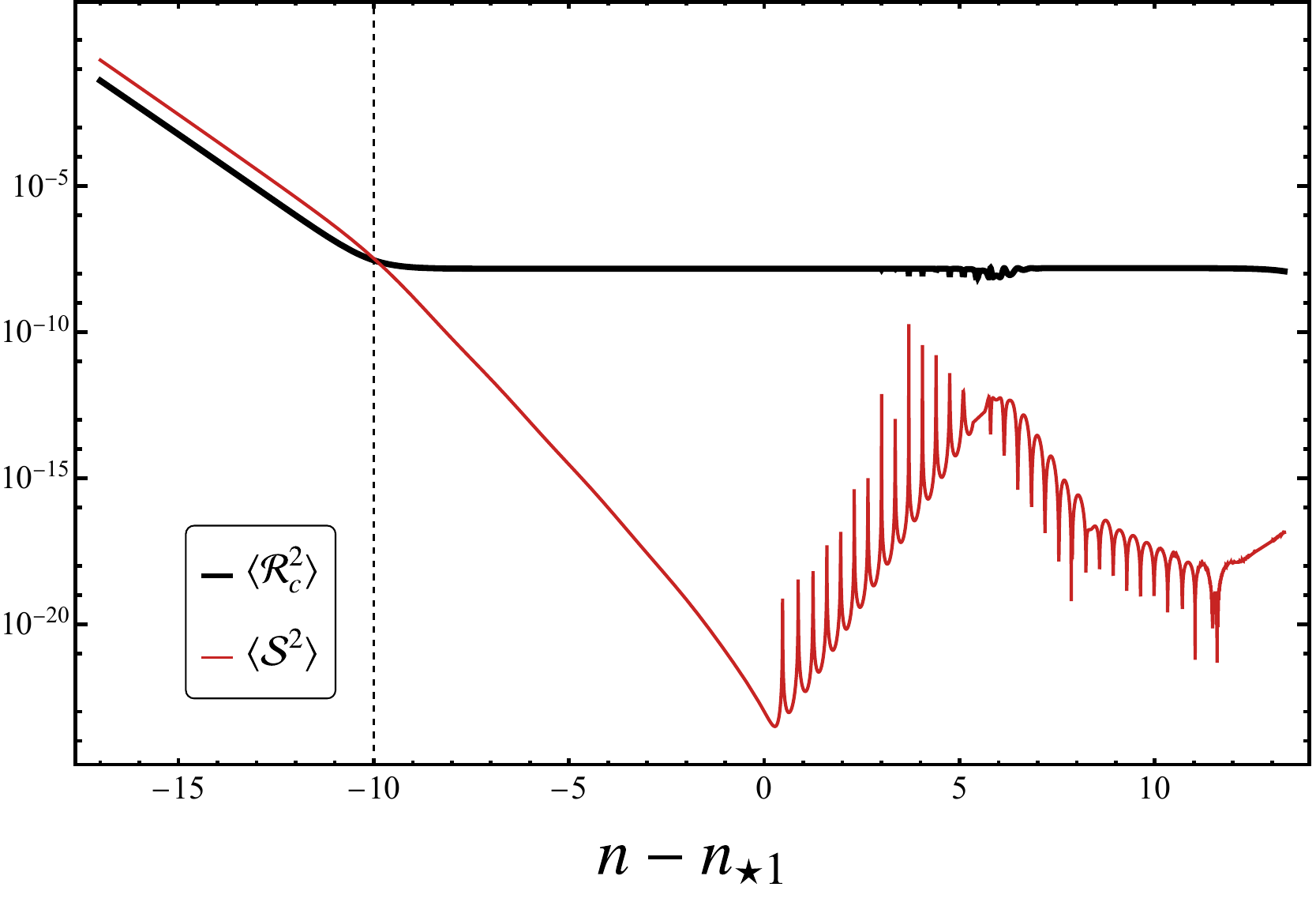}
\hspace{0.5in}
\includegraphics[width=0.7\textwidth]{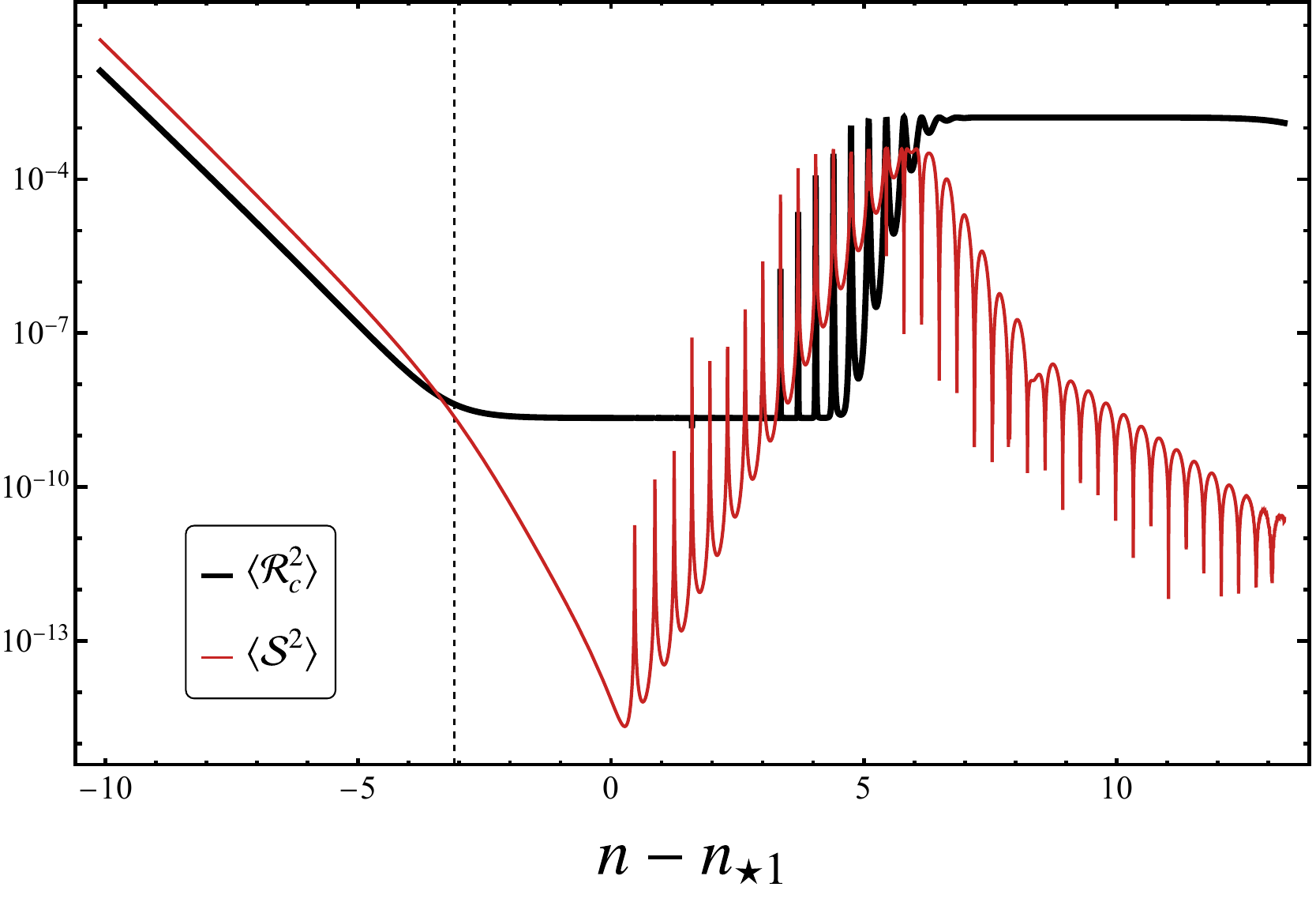}
\caption{The time evolution of the comoving curvature and non-adiabatic perturbations for specific modes. The solid black line and red line respectively show the time evolution for the two-point function of $\mathcal{R}_c$ and $\mathcal{S}$. The vertical dashed line indicates the horizon exiting time for the specified mode. The upper panel and the below panel indicate separately for $k/k_1= 10^{-4}$ and  $k/k_1= 10^{-1}$.}
\label{tevo}
\end{figure}

In the following, we analytically estimate the final amplitude of the curvature perturbation using the $\delta N$ formalism.
As we are interested in the amplitude in comoving wavenumber space, the subscript $k$ is attached to the Fourier mode, and the moment when the $k$-mode exits the horizon is denoted by $n_k$. For completeness, we also numerically solve the perturbation equations under the adiabatic vacuum initial condition, to verify our analytical approximation (See Fig.~\ref{tevo}). 

\begin{figure}[htbp]
\centering
\includegraphics[width=0.7\textwidth]{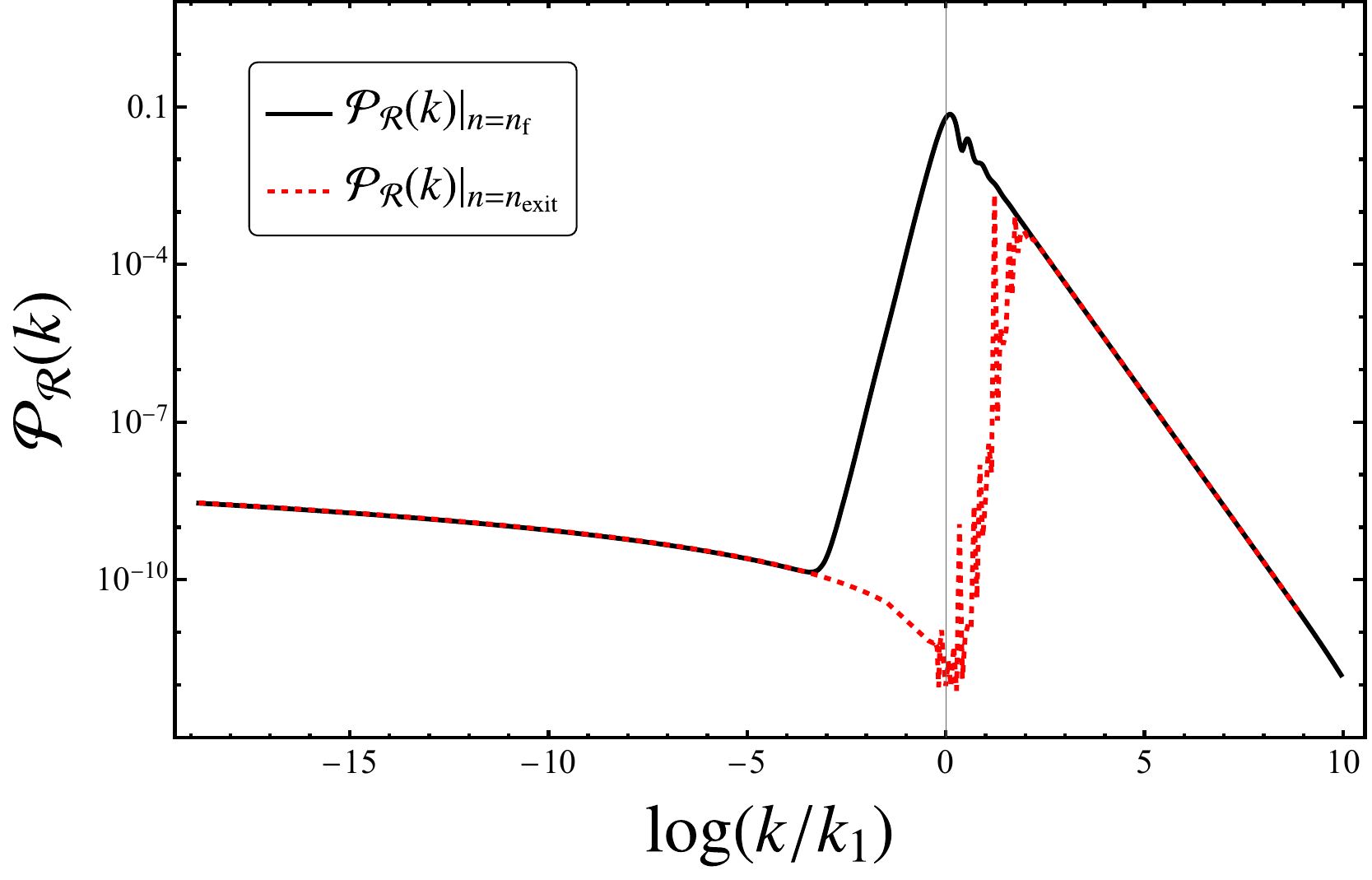}
\caption{The power spectrum numerically calculated at the horizon exiting stage and the end stage of inflation. The red dotted line indicates the curvature perturbation calculated shortly after the horizon crossing stage, i.e. $n_{\rm exit}$. In actual calculation, we set $n_{\rm exit}$ to be 2 folds after the exact horizon exiting time, which means $k=H\exp(n_{\rm exit}-2)$. The black line shows the power spectrum evaluated at the end of inflation $n_{\mathrm{f}}$, determined by the violation of slow roll condition $\epsilon_H=1$. The parameters used for this plot can be found in case 1 of Table. \ref{parameters}  }
\label{prevo}
\end{figure}


\begin{figure}[htbp]
\centering
\includegraphics[width=0.7\textwidth]{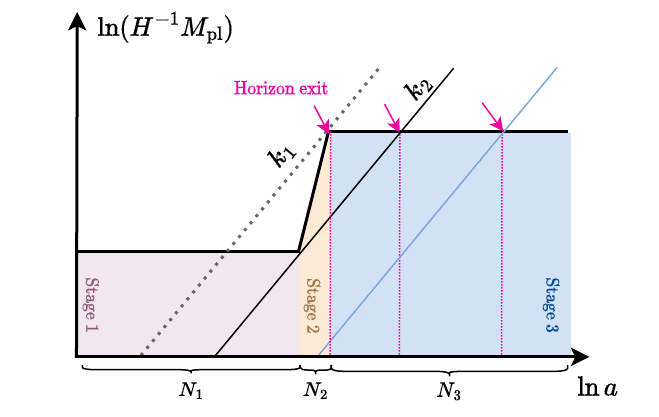}
\caption{A schematic diagram of different wave numbers and their scale evolution with respect to the Hubble horizon. Mode with $k=k_1$ is illustrated by the black dotted line, it firstly leaves the horizon before the end of St-1 and finally leaves the horizon at the beginning of St-2. Mode with $k=k_2$ is plotted by the black solid line, its scale coincides with the horizon firstly during the end of St-1, while finally exiting the horizon during early St-3. For large-scale modes with $k<k_{1}$, they exit the horizon during the first stage at $n_k<n_{k_1}$. For small-scale modes with $k>k_2$, they exit the horizon during St-3  at $n_k>n_{k_1}$. For modes with $k_1<k<k_2$,  they temporarily exit the horizon during St-1 re-enter the horizon during St-2, and make their final horizon exiting during St-3. We use the field perturbation at the last time of horizon exit (pointed out with pink arrows ) to calculate $
\delta N$. As shown on the figure, $k_1= a_{\star2} H_{\star2},k_{2}= a_{\star1} H_{\star1}$. }
\label{pertevo}
\end{figure}

\begin{figure}[htbp]
\centering
\includegraphics[width=0.7\textwidth]{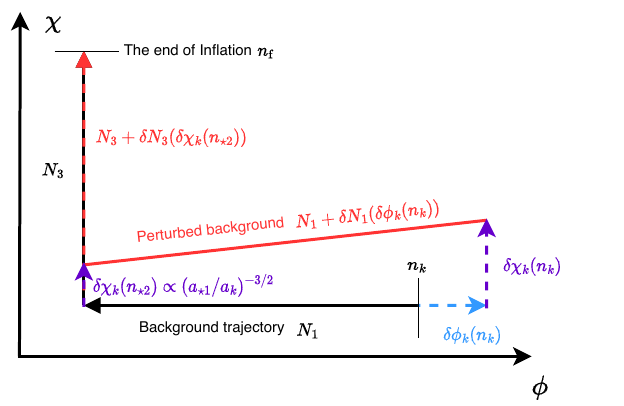}
\caption{A schematic diagram of the $\delta N$ formalism. The black line indicates the background evolution with initial condition $(\phi,\chi)$ at $n=n_k$, the pink dotted line indicates the background evolution with a perturbed initial condition $(\phi+\delta\phi,\chi+\delta\chi)$ at $n=n_{k}$. }
\label{deltaNfigure}
\end{figure}

\subsection{Perturbation equations and subhorizon evolution}
\label{perturbations}
The perturbation equations for multi-scalar fields on spatially flat slicing are expressed as \cite{Sasaki:1995aw}
\begin{align}
    &\frac{D^2\delta\psi_k^a}{dt}+3H\frac{D\delta\psi_k^a}{dt}+\frac{k^2}{a^2}{\delta\psi_k^a}+V^{;a}_{\ \ ;b}\delta\psi_k^b
    \nonumber\\
     &\qquad\qquad  -R^a{}_{bcd}\dot\psi^b\dot\psi^c\delta\psi_k^d
 -\left[\frac{1}{a^3}\frac{d}{dt}\left(\frac{a^3}{H}\dot\psi^a\dot\psi^b\right)\right]h_{bc}\delta\psi_k^b=0.
    \label{perteq}
\end{align}
We solve the above equation by rewriting it in terms of the conformal time $\eta$,
$d\eta=dt/a(t)$. Note that we have $\eta\simeq (aH)^{-1}$ during the two slow-roll stages.

During St-2, the two fields are coupled in a complicated way, which renders an analytical treatment very difficult. Fortunately, however, 
because the duration of St-2 is short, within one e-fold or so, we may focus on the modes that leave the horizon during St-1 and St-3 where the standard slow-roll approximation is valid.
Namely, let $k_1$ be the mode that leaves the horizon during St-1 and just touches the horizon at the end of St-2, and $k_2$ be the mode that touches the horizon at the end of St-1 and leaves the horizon during St-3, as illustrated in Fig.~\ref{pertevo}. We focus on the modes $k<k_1$ and $k>k_2$. 

We set $\phi^{(j)}(\eta, \mathbf{x})=\phi^{(j)}(\eta)+\delta\phi^{(j)}(\eta, \mathbf{x})$, $\chi^{(j)}(\eta, \mathbf{x})=\chi^{(j)}(\eta)+\delta\chi^{(j)}(\eta, \mathbf{x})$, respectively, where the superscript $(j)$ is for St-$j$ ($j=1,\,3$). 
The field perturbations can be quantized as
\begin{align}    
\delta\Phi_a^{(j)}(\eta, \mathbf{x})=\int \frac{d^3\mathbf{k}}{(2\pi)^{3/2}}
\left[a_\mathbf{k}u_{a}^{(j)}(\eta)e^{i\mathbf{k}\cdot\mathbf{x}}
+a^{\dagger}_{\mathbf{k}}u_{ a}^{(j)*}(\eta)e^{-i\mathbf{k}\cdot\mathbf{x}}
\right],
\end{align}
where we have introduced $(\delta\Phi_1, \delta\Phi_2)=(a\delta\phi, aF^{-1/2}\delta\chi)$.
The creation and annihilation operators satisfy the standard canonical commutation relations $\left[a_\mathbf{k}, a_{\mathbf{k'}}\right]=0$ and $[a_\mathbf{k}, a^{\dagger}_{\mathbf{k'}}]=\delta^{(3)}(\mathbf{k}-\mathbf{k'})$, and the mode functions satisfy the Klein-Gordon normalization,
\begin{align}
 u_a^{(j)}\frac{du_a^{(j)*}}{d\eta}- u_a^{(j)*}\frac{du_a^{(j)}}{d\eta}=i\,,
\end{align}
in the asymptotic past, $\eta\to-\infty$.
To the first slow-roll order, the equations of motion for $u_a^{j}$ are expressed as
\begin{align}
    \frac{d^2u_a^{(j)}}{d\eta^2}+k^2u_a^{(j)}=\frac{1}{\eta^2}\left(2u_a^{(j)}+3\epsilon_a{}^bu_b^{(j)}\right),
    \label{ueom}
\end{align}
where
\begin{align} 
\epsilon_{a}{}^b&\equiv\epsilon_H\delta_{a}^{b}+\left(\delta^{b}_{d}h_{ac}-\frac{1}{3}M_{\mathrm{pl}}^2R^{b}{}_{acd}\right)\frac{\dot\psi^c\dot\psi^d}{M_{\mathrm{pl}}^2H^2}-\frac{U_{;a}{}^{;b}}{3H^2},\label{epsilonab}\\
U_{;a}{}^{;b}&\equiv h^{bc}\frac{D}{d\psi^c}\left(\frac{\partial U}{\partial \psi^a}\right).\label{Uab}
\end{align}
The explicit expressions for $R^{b}{}_{acd}$ are given in \eqref{fieldcompo}.
The slow-roll parameter is given by $\epsilon_H= h_{ab}\dot\psi^a\dot\psi^b/(2H^2M_{\mathrm{pl}}^2)$.
Since the first two terms in \eqref{epsilonab} is proportional to $\epsilon_H$ if we ignore them and retain only the terms proportional to the potential term, $\epsilon_a{}^b$ becomes diagonal,
\begin{align}    \epsilon_{a}^b&\approx -\delta_a^b\eta_a\,;
\quad \eta_1\equiv \frac{M_{\rm pl}^2U_{;\phi}{}^{;\phi}}{U}\,,
~\eta_2\equiv\frac{M_{\rm pl}^2U_{;\chi}{}^{;\chi}}{U}\,.
\end{align}
This implies that $\delta\phi^{(j)}$ and $\delta\chi^{(j)}$ are decoupled from each other during St-1 and St-3. The explicit expressions for $\epsilon_a{}^{b(j)}$ ($j=1,\,3$) are
\begin{align}
&\epsilon_\phi^{\phi(1)}\approx\frac{4}{3F(\phi(n_k))}\,,
\quad\epsilon_\chi^{\chi(1)}\approx-4\xi\,,
\\
&\epsilon_\phi^{\phi(3)}\approx-\frac{4}{3}\mu^2\,,
\quad\epsilon_\chi^{\chi(3)}\approx4(A-1)\xi\,.
\end{align}

With the adiabatic vacuum initial condition, $u_a^{(1)}\propto e^{-ik\eta}$ as $k\eta\to-\infty$, \eqref{ueom} can be solved as
\begin{align}
    u_a(\eta)
    =\frac{\sqrt{\pi}}{2\sqrt{k}}\sqrt{-k\eta}\,e^{i\nu_a\pi/2}H_{\nu_a}^{(1)}(-k\eta)\,;
\quad
\nu_a=\sqrt{\frac{9}{4}-3\eta_a}\,,
\end{align}
where $H_{\nu}^{(1)}(z)$ is the Hankel functions of the first kind, and we have omitted the superscript $(j)$ that distinguishes between St-1 and St-3.
In the case when $\nu_a$ becomes imaginary, we set $\nu_a=i\nu_a'$ where $\nu_a'>0$. 
We also note that we have assumed $k\gg k_2$ for the mode functions $u_a$ at St-3. Rigorously speaking they should be obtained by solving them through St-1 and St-2 with the adiabatic vacuum initial condition at St-1. Nevertheless, for those modes that are still deep inside the horizon at $n=n_3$, their behavior will not be affected by the transitions. Thus we may apply the adiabatic vacuum initial condition for the mode functions at St-3 as well.

Using the solutions obtained above, we can immediately compute the power spectrum of the fields $\delta\psi^a$,
\begin{align}
\langle\delta\psi^{a}(\mathbf{k})\delta\psi^{\dagger b}(\mathbf{k}')\rangle
  &=h^{ab}(n)P_{\nu_a}(k,\eta)\delta^3(\mathbf{k}-\mathbf{k}')\,;
\nonumber\\
P_{\nu_a}(k,\eta)
&=\frac{H(n)^2}{2k^3}\frac{\pi}{2}|e^{i\nu_a\pi}|\left|H_{\nu_a}^{(1)}(-k\eta)\right|^2\,,
    \label{corfunc}
\end{align}
where $n$ is the time in terms of the number of e-folds, $dn=Ha\,d\eta$. 
On superhorizon scales $-k\eta\ll1$, however, the above result becomes unreliable because the non-trivial evolution of the background invalidates the pure de Sitter space approximation. 
This means it is reliable only up to a few e-folds after the mode crosses the horizon.
On the other hand, on superhorizon scales, the so-called separate universe approach becomes valid and hence the evolution of the field fluctuations can be well approximated by the perturbation of the classical background trajectory.
Thus, to obtain reasonable estimates of $P_{\nu_a}$ on superhorizon scales, one has to match the solution valid inside the horizon with that valid outside the horizon. A fairly accurate way to do this is to retain only the terms that dominate at $-k\eta\ll1$ in \eqref{corfunc} and set $-k\eta=1$ in the final expression. This gives a fairly accurate estimate of $P_{\nu_a}$ that may be used as the initial conditions for the superhorizon evolution. We call the values of $P_{\nu_a}$ estimated in this manner the horizon crossing values, which we denote by $P_{\nu_a}(n_k)$.

The horizon crossing values of the power spectrum for the field fluctuations on flat slicing are estimated as
\begin{align}
    &{\cal P}_\phi(k,n_k)\equiv\frac{k^3}{2\pi^2}P_{\nu_\phi}(n_k)
    \approx \dfrac{H_k^2}{(2\pi)^2} \qquad\mathrm{for}\ k<k_1\,,\\
    ~\nonumber\\
    &{\cal P}_\chi(k,n_k)\equiv\frac{k^3}{2\pi^2}F(n_k)P_{\nu_\chi}(n_k)\approx
   \left\{ \begin{array}{ll}
     {g_1^2}{F_k}\dfrac{H_k^2}{(2\pi)^2}&\quad\mathrm{for}\ k<k_1\,,
     \\
     ~\\
     g_2^2\dfrac{H_k^2}{(2\pi)^2}&\quad\mathrm{for}\ k>k_2\,, 
    \end{array}
    \right.
    \label{pertsolu}
\end{align}
where $F_k=F(n_k)$ and $H_k=H(n_k)$, and ${\cal P}_a(k,n)$ ($a=\phi$ and $\chi$) are the fluctuation spectra having the scalar field dimensions which we sometimes denote by $\langle\delta\phi_k^2(n)\rangle$ and $\langle\delta\chi_k^2(n)\rangle$ for $\phi$ and $\chi$, respectively.
In the above, we have not bothered to compute the spectrum of $\delta\phi$ at St-3 as it does not contribute to the final curvature perturbation spectrum.  
where $F(n_k)\gg 1$ except for the stage near the end of St-1, and 
\begin{align}
    g_{1}^2&\equiv
      \frac{\pi}{2}e^{-\pi \sqrt{12\xi-\frac94} }
      \left|\frac{\Gamma\left(i{\sqrt{12\xi-\frac94}}\right)}{i\pi}
      +\frac{1+\coth (\pi\sqrt{12\xi-\frac94})}{\Gamma\left({1+i\sqrt{12\xi-\frac94}}\right)}\right|^2
      \,,\\
    g_{2}^2&\equiv\frac{2^{\sqrt{9+48\xi(A-1)}-1}}{\pi}\left|\Gamma\left(\sqrt{\frac94+12\xi(A-1)}\right)\right|^2.\label{g2def}
\end{align}
The above expressions for $g_1^2$ and $g_2^2$ are rather complicated. However, under the assumptions that $12\xi\gg1$ and $12\xi(A-1)\ll1$, they reduce to $g_1^2\approx(12\xi)^{-1/2}$ and $g_2^2\approx1$, which are crude but qualitatively useful approximations.

\subsection{Superhorizon evolution and \texorpdfstring{$\delta N$}{Lg}}
\label{superhorizon}
Now let us consider the evolution of the mode functions on superhorizon scales.
Except for the era near the end of St-1 and during St-2, the evolution is essentially trivial as the system resembles that of single-field slow-roll inflation. On the other hand, the evolution during and just before the beginning of St-2 needs a careful analysis. In order to take into account such a complicated case, the $\delta N$ formalism is most convenient. In this formalism, the evolution of the field fluctuations is taken into account by considering a family of classical (ie, homogeneous and isotropic) background universes, and the final comoving curvature perturbation at the end of inflation is given by the difference in the number of e-folds between the perturbed trajectory and the fiducial trajectory, caused by the difference in the initial condition due to the field fluctuations at horizon crossing.  To be specific, the $\delta N$ formalism tells us that
\begin{align}
   {\cal R}(n_f)&=\delta N(n_k;n_f)\equiv N(\psi^a(n_k)+\delta\psi^a(n_k);n_f)-N(\psi^a(n_k);n_f)
\nonumber\\
&=\left.\frac{\partial N}{\partial \psi^a}\right|_{n_k}\delta\psi^a(n_k)
+O\left((\delta\psi^a)^2\right)\,,
\end{align}
where $N(\psi^a(n_k);n_f)$ is the number of e-folds from the horizon crossing time of the comoving scale $k$ until the end of inflation, $\psi^a$ represent all the dynamical degrees of freedom that contribute to the background evolution, and $\delta\psi^a$ are the fluctuations evaluated on the flat slicing.

In this section, we derive the final curvature perturbation both numerically and analytically. Numerically we solve the equations for the field fluctuations \eqref{perteq} on flat slicing exactly without any approximations from the stage deep inside the horizon until the end of inflation. 
An advantage of this method is, in addition to its quantitative accuracy, one can follow the evolution of the field fluctuations as well as that of the comoving curvature and isocurvature perturbations throughout the entire stage, which helps us to understand the two-field dynamics better.
Analytically, using the result obtained in the previous subsection as the initial condition for the superhorizon evolution, we resort to the $\delta N$ formalism to evaluate the final curvature perturbation.


To compute $\delta N$, it is convenient to divide $N$ into three, $N=N_1+N_2+N_3$, where $N_j$ is the contribution from St-$j$ ($j=1,2,3$). 
The expressions for $N_1$, $N_2$ and $N_3$ are given by \eqref{N1}, \eqref{N2} and \eqref{ends3}, respectively.
When we divide the contributions, it is important to take into account the field perturbations at the matching epochs. Thus for the modes that leave the horizon during St-1, for example, we have
\begin{align}
&N(\psi^a(n_k);n_f)
\nonumber\\
&\quad=N_1\bigl(\psi^a(n_k);\psi^a(n_{\star 1})\bigr)
+N_2\bigl(\psi^a(n_{\star 1});\psi^a(n_{\star 2})\bigr)+N_3\bigl(\psi^a(n_{\star 2});\psi^a(n_f)\bigr)\,,
\end{align}
and we have to take into account the perturbations of $\psi^a(n_{\star 1})$ and $\psi^a(n_{\star 2})$
at the two boundaries.

Let us first consider the modes that leave the horizon during St-1.
The contribution from St-1 takes the form,
\begin{align}\label{N1app}
\delta N_1\equiv\delta(n_{\star1}-n_k)&=N_1\left[\left(\phi+\delta\phi, \chi+\delta\chi\right)_{n_k}; \left(\phi+\delta\phi, \chi+\delta\chi\right)_{n_{\star 1}}\right]\nonumber\\
&\qquad-N_1\left[\left(\phi, \chi\right)_{n_k}; \left(\phi, \chi\right)_{n_{\star 1}}\right]\nonumber\\
&\approx\left(\frac{\partial N_1}{\partial\phi}\delta\phi+\frac{\partial N_1}{\partial\chi}\delta\chi\right)_{n_k}
+\left(\frac{\partial N_1}{\partial\phi}\delta\phi+\frac{\partial N_1}{\partial\chi}\delta\chi\right)_{n_{\star 1}}
\end{align}
During St-1, $\delta\chi$ decays as $\delta\chi\propto a^{-3/2}$, so neither $\delta\chi(n_k)$ or $\delta\chi_{n_{\star 1}}$ contributes to $\delta N_1$. As for $\delta\phi$, since the end of St-1 is defined by the value of $\phi$, it automatically follows that $\delta\phi|_{n_{\star 1}}=0$. 
Hence, we end up with
\begin{align}\label{N1appfin}
\delta N_1\approx\left(\frac{\partial N_1}{\partial\phi}\delta\phi
\right)_{n_k}\quad\mathrm{for}\ k< k_1\,.
\end{align}
This result can be immediately derived from the expression for $N_1$ given by \eqref{N1}. The number of e-folds is independent of $\chi$ due to the single-field slow-roll nature of St-1.

About the contribution from St-2, as we can see from the expression for $N_2$ given by \eqref{N2}, its perturbation vanishes. This is of course an approximation. But because of the shortness of St-2, $N_2=O(1)$, its contribution can be neglected in comparison with the contributions from St-1 and St-3. Hence we have 
\begin{align}
    \delta N_2=0\,.
    \label{dN2}
\end{align}

As for the contribution from St-3, since the end of inflation is determined by fixed values of $\phi$ and $\chi$, the only contribution is from the boundary at $n=n_{\star 2}$. Thus
\begin{align}
\delta N_3\approx
  \left(\frac{\partial N_3}{\partial\phi}\delta\phi+\frac{\partial N_3}{\partial\chi}\delta\chi\right)_{n_{\star 2}}
  =\left(\frac{\partial N_3}{\partial\chi}\delta\chi\right)_{n_{\star 2}}\,,
  \label{dN3}
\end{align}
where we have used the fact that St-3 is again an effectively single-field slow-roll stage with $\chi$ being the inflaton. Thus $\delta N_3$ is determined solely by $\delta\chi_{n_{\star 2}}$. To determine it, we have to solve the superhorizon evolution of $\delta\chi$ from $n=n_k$ to $n=n_{\star2}$.
This can be easily obtained by perturbing the background solution \eqref{chisol1}. We find
\begin{align}
\langle\delta\chi^2(n_{\star2})\rangle
\approx\langle\delta\chi^2(n_{\star1})\rangle
=\langle\delta\chi^2(n_k)\rangle\exp[-3(n_{\star1}-n_k)]\frac{(F/H)_{\star1}}{(F/H)_{k}}\,,
\end{align}
where $(FH)_k=(FH)(n_k)$ and we have assumed $\langle\delta\chi^2(n_k)\rangle\approx\langle\delta\chi'{}^2(n_k)\rangle/\omega^2$ on average, and ignored the evolution of $\delta\chi$ during St-2. 
Replacing $\langle\delta\chi^2(n_k)\rangle$ by $\langle\delta\chi_k^2(n_k)\rangle={\cal P}_\chi(k,n_k)$ obtained in the previous subsection,
and noting that $k/k_1=e^{-(n_{\star 1}-n_k)}$, we obtain
\begin{align}
\langle\delta\chi_k^2(n_{\star 2})\rangle
=\left(\frac{k}{k_1}\right)^3\langle\delta\chi_k^2(n_k)\rangle\approx
\left(\frac{k}{k_1}\right)^3\frac{F_{\star2}H_k^3}{H_{\star 2}(2\pi)^2}\,,
\label{dchi2}
\end{align}
where \eqref{phiN1} is used to express $F(n_k)$ in terms of $k/k_1$. Thus apart from the small $\ln k/k_1$ correction, the spectrum shows a rapid increase $\propto k^3$ up to $k=k_1$. The above gives the initial condition of $\delta\chi$ at $n=n_{\star 2}$ for the evaluation of $\delta N_3$ in \eqref{dN3}.

As we stated before, it is analytically formidable to consider the modes that exit the horizon during St-2. However, fortunately, since the duration of St-2 is short, $N_2\sim1$, ignoring these modes does not change the qualitative picture. This is also confirmed by numerical computations. Thus we now turn to the modes that leave the horizon during St-3, ie, $k>k_2$. Apparently $N=N_3$ for these modes. In addition, we have seen that the single-field slow-roll approximation is accurately valid at St-3. 
Therefore, it is only necessary to consider the initial fluctuation of $\chi$ when the mode $k$ exits the horizon, $n=n_k$. 
The number of e-folds from the horizon crossing time $n=n_k$ till the end of inflation $n=n_f$ is given by \eqref{ends3} with the replacement of $n_{\star 2}$ by $n_k$. Taking the variation of it at $n=n_k$ gives
\begin{align}\label{N3appshort}
\delta N(n_k,n_f)=\delta N_3\approx\left(\frac{\partial N_3}{\partial\chi}\delta\chi\right)_{n_{k}},\quad\mathrm{for}\ k> k_2.
\end{align}

Combining \eqref{N1appfin}, \eqref{dN3} and \eqref{N3appshort}, the primordial curvature perturbation at the end of inflation is approximately given as
\begin{equation}
\begin{aligned}
 \mathcal{R}(n_k,n_{\mathrm{f}})
&\approx\left\{ 
\begin{aligned}
&\left(\frac{\partial N_1}{\partial\phi}\delta\phi
\right)_{n_k}
+\left(\frac{\partial N_3}{\partial\chi}\delta\chi\right)_{n_{\star 2}}&\quad\mathrm{for}\ k<k_1\,,\\
&\left(\frac{\partial N_3}{\partial\chi}\delta\chi\right)_{n_{k}}&\quad\mathrm{for}\ k> k_2\,.
\end{aligned}
\right.
\end{aligned}
\label{deltaN}
\end{equation}
We note that $\delta\phi(n_k)$ and $\delta\chi(n_k)$ are the horizon crossing values of the $k$-mode scalar perturbations, while $\delta\chi(n_{\star2})$ is the value at $n=n_{\star2}$ on superhorizon scales determined by the perturbed background evolution after horizon crossing with the initial condition give by $\delta\chi(n_k)$.

To summarize, using the expressions for $N_1$ in \eqref{N1} and $N_3$ in \eqref{ends3} (with $n_{\star 2}$ replaced by $n_k$), together with the perturbation amplitude at horizon crossing \eqref{pertsolu} as well as $\delta\chi(n_{2*})$ given by \eqref{dchi2},
we obtain an approximate dimensionless power spectrum of the comoving curvature perturbation at the end of inflation as
\begin{equation}
\mathcal{P}_{\mathcal{R}}(k)\approx\frac{M^2}{16\pi^2 M_{\rm pl}^2}\times\left\{ 
 \begin{aligned}
 &\quad\left[\frac{2}{3}\left(\ln\frac{k_1}{k}+\frac{3}{4}F_{\star1}\right)^2
 +g_1^2h^2\left(\frac{M_{\rm pl}}{\chi_0}\right)^{2} \frac{H_{\rm exit}}{H_{\star2}}
 \left(\frac{k}{k_1}\right)^{3}\right]\\
    &\qquad\times\left[1-\left(\frac{4}{3}\ln\frac{k_1}{k}+F_{\star1}\right)^{-1}\right]^2
    \quad\mathrm{for}\ k<k_1\,,\\
 &\quad\frac{g_2^2h^2}{\mu^{2}}\left(\frac{M_{\rm pl}}{\chi_0}\right)^{2} \left(\frac{k}{k_2}\right)^{-\beta}\quad\mathrm{for}\ k>k_1\,,
    \end{aligned}
    \label{prform}
    \right.
\end{equation}
where we have used $H^2=M^2(1-F^{-1})^2/4$ during St-1, $H^2=V_0/(3M_{\mathrm{pl}})^2$ during St-3, $F_{\star2}=1$, and introduced
\begin{align}
    &h\equiv\frac{2(A-1)}{\beta A}\frac{\chi_0^2}{(\chi_0+(A-1)\chi_{\star2})^2}\simeq\frac{2(A-1)}{\beta A},
    \label{hdef}
\end{align}
An intriguing feature of the above spectrum is that the steep growth is proportional to $k^3$ for $k<k_1$. This arises from the fact that $\chi$ is heavy and its mean square fluctuations are damped in proportion to $a^{-3}$, where ``3" comes from our spatial dimensions. This feature is expected to be quite universal in many two-stage models in which one of the fields that governs the later stage of inflaton is heavy during the first stage. In this respect, an additional point to be mentioned is that, although we do not discuss the case when $\chi$ is relatively light so that it undergoes over-damping instead of under-damping, one can easily guess that the growth would become shallower, as the mean square fluctuations would not decrease as fast as $a^{-3}$.
Another feature is the spectral index $-\beta$ at $k>k_2$, where $\beta>0$. This is actually a common feature in single-field slow-roll inflation with a concave potential, provided that the contribution from the slow-roll parameter $\epsilon_H$ is negligibly small.
\begin{figure}[htbp]
\centering
\includegraphics[width=0.7\textwidth]{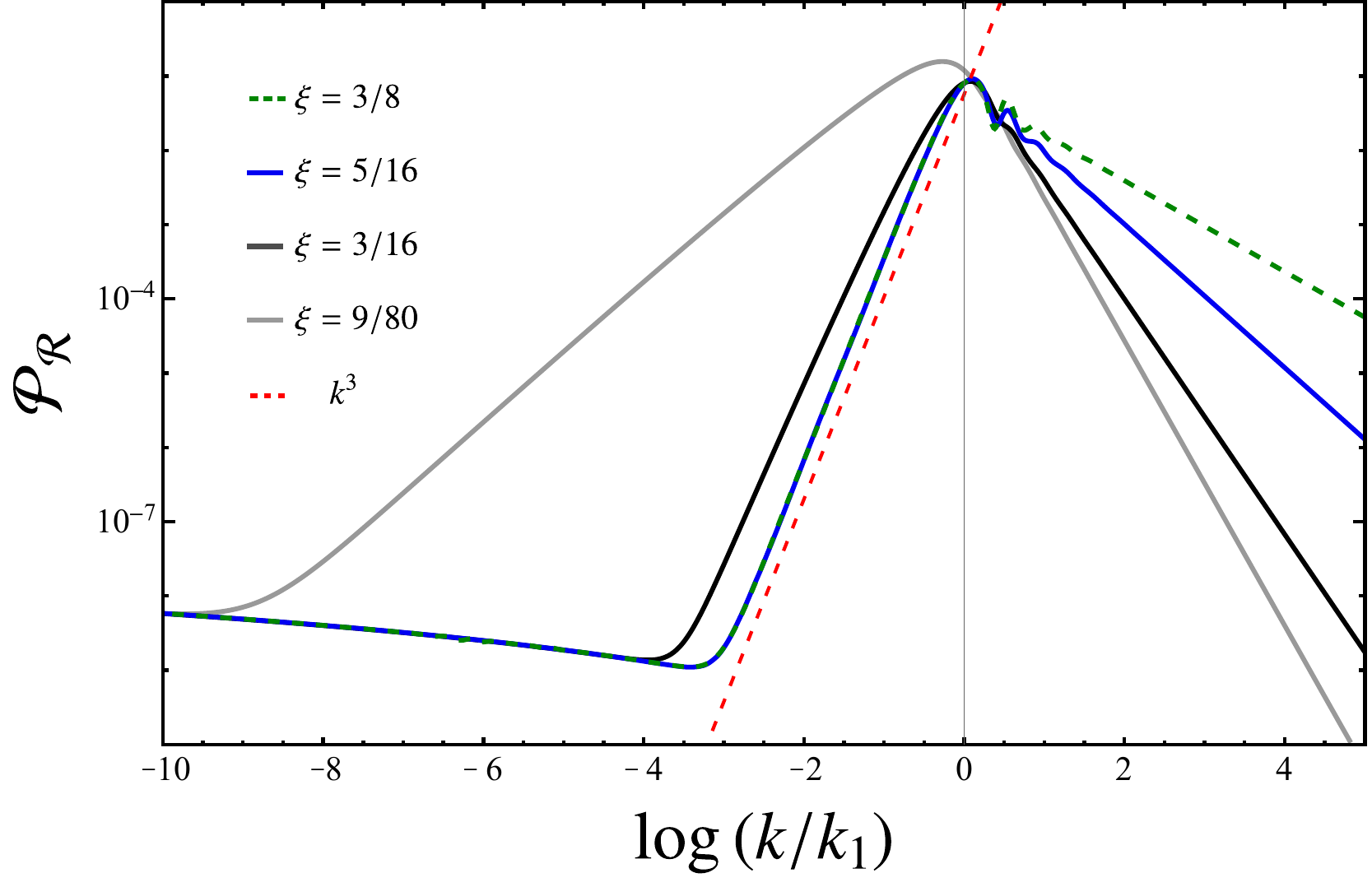}
\caption{The numerical result of the power spectrum for different $\xi$ and the same $\mu$.
 For $\xi<3/16$, the growth rate of $ \mathcal{P}_{\mathcal{R}}(k)$ is determined by $\xi$ and slower than $k^3$. For $\xi\geq3/16$,  the growth of the power spectrum is $\sim k^3$, which means the slope is parameter independent. The parameters $m, M$ and $\chi_0$ used for numerical calculation can be found in case 1 of Table. \ref{parameters}, we specifically choose $A\xi=15/32$ as a constant.}
\label{growthrate}
\end{figure}

\begin{figure}[htbp]
\centering
\includegraphics[width=0.7\textwidth]{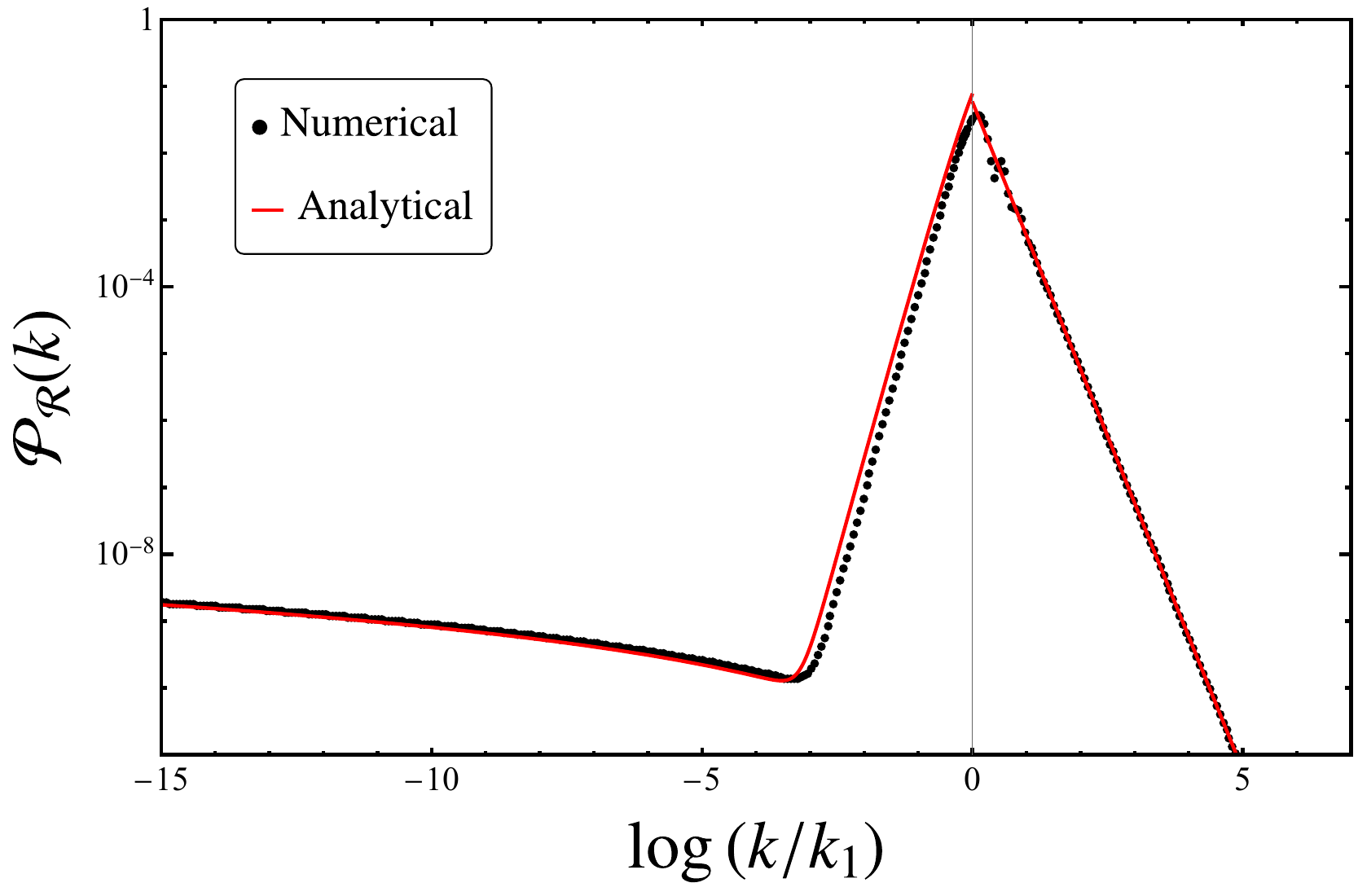}
\caption{The numerical power spectrum and its analytical approximation. The black points are numerical results calculated by solving the perturbation equations, the red line is the analytic approximation given by Eq.~\eqref{prform}. The parameters are supplied as case 1 Table \ref{parameters}. The broken power-law feature given by Eq.~\eqref{prform} fits the numerical result well. The parameters used for calculation can be found in case 1 of Table. \ref{parameters}.}
\label{pr}
\end{figure}

In Fig.~\ref{pr}, we plot the power spectrum of primordial power spectrum both analytically and numerically, with the identification of $k_1=k_2$ for the analytical expression \eqref{prform}. There appears discontinuity in the analytical result at $k=k_1$ (see the red curve of Fig.~\ref{pr}) due to the neglection of the contribution from St-2. However, from the comparison with numerical calculations (see dotted curve), it is obvious that the deviation is small, which confirms our expectation that the contribution from St-2 can be safely neglected.
We also note that the peak value of the power spectrum at $k=k_1$ is
\begin{equation}
    \begin{aligned}
    \mathcal{P}_{\mathcal{R}}^{\mathrm{peak}}\approx\frac{g_2^2h^2}{16\pi^2\mu^{2}}\left(\frac{M}{\chi_0}\right)^{2}=\frac{g_2^2h^2}{12\pi^2}\frac{V_0}{M^2_{\mathrm{pl}}\chi^2_0}\,,
    \label{prpeak}
\end{aligned}
\end{equation}
where $g_2$ and $h$ are defined in \eqref{g2def} and \eqref{hdef}, respectively. 
It may be useful to consider the peak amplitude in the limit $|\xi(A-1)|\ll1$. In this limit, it reduces to 
\begin{align}
 \mathcal{P}_{\mathcal{R}}^{\mathrm{peak}}\approx\frac{4}{3\pi^2}\frac{M_{\rm pl}^2V_0}{\chi_0^2\chi_g^4}
=\frac{A\xi}{\pi^2}\frac{m^2}{\chi_0^2}\,.
\end{align}
 
Before concluding this section, let us discuss our model in light of the observational constraints.
It is known that the observational data of Cosmic Microwave Background (CMB) anisotropies gives strong constraints on the scalaron mass $M$. 
On scales $k\ll k_1$, the power spectrum of the comoving curvature perturbation is given by
\begin{align}\label{prir}
 \mathcal{P}_{\mathcal{R}}(k\ll k_1)\approx\frac{3M^2}{128\pi^2}F^2(n_k)\,,
\end{align}
with the spectral index,
\begin{align}
    n_s\equiv1+\frac{d\ln\mathcal{P}_{\mathcal{R}}}{d\ln k}=1-\frac{8}{3}\frac{1}{F(n_k)}\,.
    \label{nsr2}
\end{align}
The amplitude on the CMB scale is constrained as $\mathcal{P}_{\mathcal{R}}\approx2\times10^{-9}$. 
Inserting this value into \eqref{prir}, we obtain $M\simeq2\times10^{-5}M_{\mathrm{pl}}$.

To evaluate the spectral index, we resort to the space-time diagram Fig.~\ref{pertevo}, from which we see that
\begin{align}\label{Nkrelation}
n_{\star 1}-n=\ln \frac{k_2}{k}.
\end{align}
Hence, using the solution for $F$ given by \eqref{phiN1}, we obtain
\begin{align}
     n_s(k)=1-\frac{8}{3}\frac{1}{F(k_{\rm CMB})}=1-\frac{8}{3}\left(\frac{4}{3}\ln\frac{k_1}{k}+\frac{4}{3}\ln\frac{k_2}{k_1}+F_{\star1}\right)^{-1},
\end{align}
where $F_{\star1}\approx1+2/\sqrt{3}$ as given in \eqref{Fstar1} and $\ln(k_2/k_1)=\ln\mu-N_2\simeq(1/3)\ln\mu$. 
It is obvious that $n_s$ monotonically approaches unity as $k/k_1\to0$. For our interest, $k_1/k_{\mathrm{CMB}}\lesssim10^{17}$ (which corresponds to the mass of PBHs larger than $10^{-16} M_\odot$).
Thus for reasonable values of $\mu^2$, $F(k_{\mathrm{CMB}})=(4/3)\ln(k_1/k_{\rm CMB})\lesssim60$. 
In comparison with the observed value $n_s(k_{\mathrm{CMB}})\simeq 0.965$, the above gives $n_s$ too small. 
In other words, we need $\mu\gtrsim10^{21}$ if we are to fit the data, which makes St-2 too long and the energy scale of St-3 too low. This means our model in its original form is not a viable model.

As discussed in~\cite{Pi:2017gih}, this situation can be remedied by adding a small correction term proportional to $R^3$ to the action, 
\begin{align}    f(\chi,R)=\left(R+\frac{R^2}{6M^2}+q\frac{R^3}{3M^4}\right)
-\frac{1}{M_\mathrm{pl}^2}\xi R(\chi-\chi_0)^2,
\end{align}
where $|q|\ll1$. The transformation to the Einstein frame gives the effective potential,
\begin{align}
    U^{R^3}(\phi,\chi)=&\frac{3}{4}M^2M_\mathrm{pl}^2W^2(\phi)\bigl[1-6qF(\phi)W(\phi)\bigr]
   +\frac{V(\chi)}{F^2(\phi)}+O(q^2)\,,
\end{align}
where $W(\phi)$ is defined in \eqref{Wdef}.

\begin{figure}[t]
\centering

\includegraphics[width=0.7\textwidth]{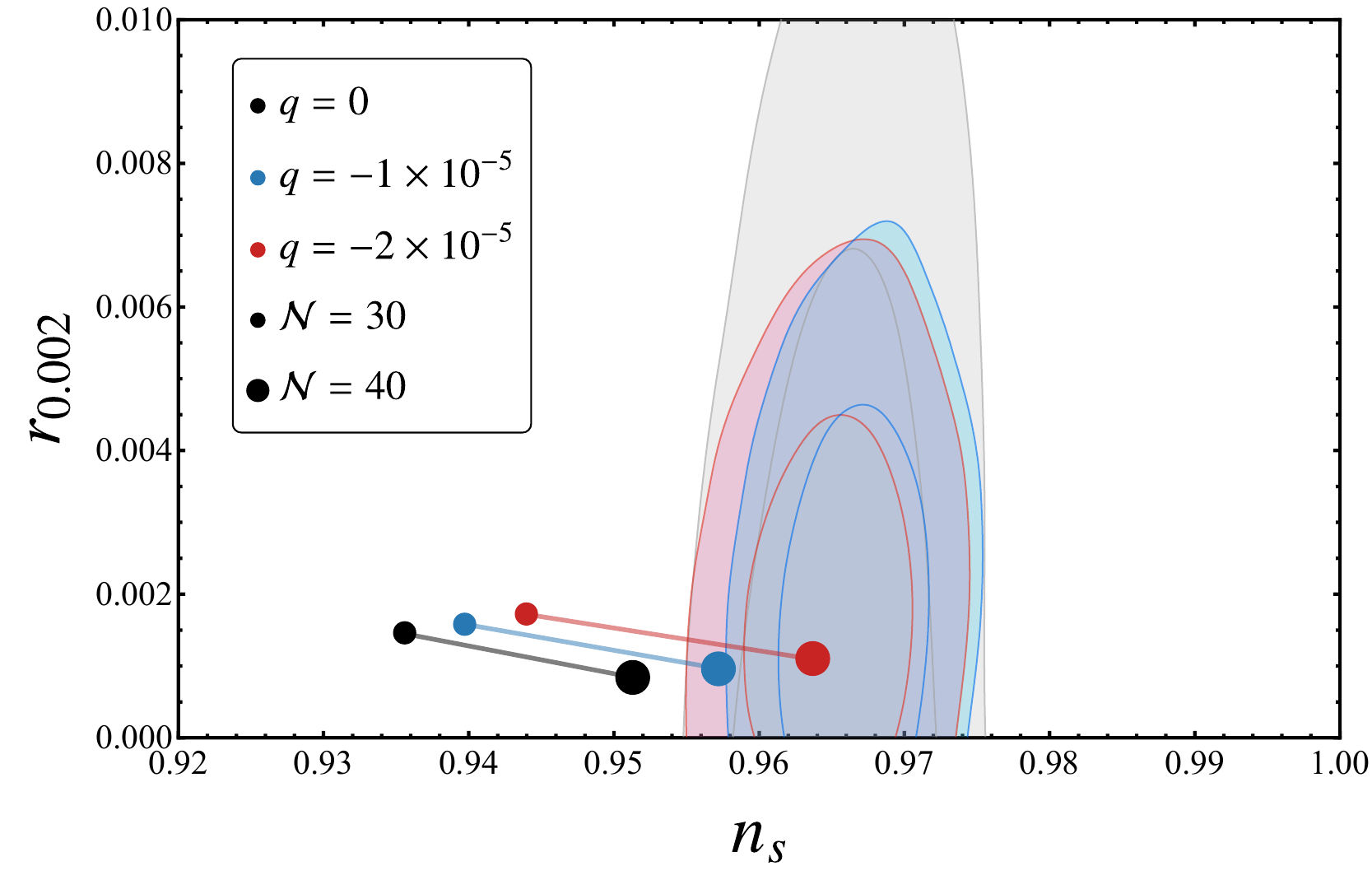}
\caption{The $n_s-r$ plot, showing its dependence on the e-folding number of St-1 from $n_\mathrm{CMB}$ and $q$. In this figure, $ \mathcal{N}=n_{\star1}-n_{\mathrm{CMB}}$, counting for the e-folding number from $n_\mathrm{CMB}$ to the end of St-1.  }
\label{nsr}
\end{figure}

Since St-1 can be approximated as single field inflation driven by the scalaron $\phi$, the spectral index is easily estimated by the formula, 
\begin{align}\label{ns}
    n_s-1\approx2\eta_V-6\epsilon_V   
\end{align}
where the slow-roll parameters are defined by the effective potential, $\epsilon_V\equiv M_{\mathrm{p}}^2(U_{,\phi}/U)^2/2$ and $\eta_V\equiv M_{\mathrm{pl}}^{2}U_{,\phi\phi}/U$. 
In the regime $|qF|\ll1$, we have
\begin{align}
  \epsilon_V^{q}\approx\frac{4}{3}\left[\frac{F-1-3qF^3}{(F-1)^2}\right]^2,\qquad\eta_V^{q}\approx-\frac{4}{3}\frac{F-2+3qF^3}{(F-1)^2}\,,\label{etaR3}
\end{align}
where the superscript $q$ denotes that they are for the potential with the $qR^3$ correction.
Thus for $F(k_{\rm CMB})\gg1$ and $|qF^2(k_{\mathrm{CMB}})|\ll1$, we obtain
\begin{align}
{n_s^{q}(k_{\mathrm{CMB}})\approx{n_s}(k_{\mathrm{CMB}})}-8qF(k_{\mathrm{CMB}})\,,
\end{align}
where $n_s(k_{\rm CMB})$ is the spectral index for the original $R^2$ model, and we have neglected the contribution from $\epsilon_V$ since we have $\epsilon_V=O(|\eta_V|/F)$ for $F\gg1$. 
Since we have $F(k_{\mathrm{CMB}})\sim50$, we can obtain the observed value $n_s^{q}(k_{\mathrm{CMB}})\sim0.965$ for $q\sim -10^{-5}$. 
Numerical results shown in Fig.~\ref{nsr} confirm this conclusion.

Finally, we note that this small correction term does not affect the amplitude of the curvature perturbation.
It is easy to see that
\begin{align}
    \frac{\mathcal{P}^{q}_{\mathcal{R}}(k)}{\mathcal{P}_{\mathcal{R}}(k)}\approx 1+6F^2q+\mathcal{O}(q^2)\quad\mathrm{for}\quad |F^2q|\ll1\,,
    \label{prratio}
\end{align}
which implies that there is virtually no effect of the $qR^3$ correction on the amplitude of the power spectrum. 

\section{PBHs and induced GWs}
In this section, we study the PBH formation and induced GWs from the peak of the curvature perturbation we obtained. First, we consider the PBH formation based on the simple Press-Schechter formalism. Then we consider the induced GWs. We also mention the possibility of explaining the recently reported evidence of a GW background by pulsar timing array (PTA) collaborations~\cite{Reardon:2023gzh, Antoniadis:2023rey, Xu:2023wog} by our model.
\label{appliance}
\subsection{Primordial black hole formation}
We first discuss the PBH formation from the curvature perturbation assuming the Gaussian probability distribution in the Press-Schechter formalism. In this formalism, we assume that the PBH formation criterion is determined by a threshold in the density contrast $\delta\equiv(\rho-\bar\rho)/\bar\rho$ defined on the comoving slicing.  
Although there have been many discussions and new proposals about how to improve this simplest criterion, since it is not our purpose to dwell into this issue, we take this simplest approach. In this sense, one should not expect any definite, quantitative conclusions from the discussion below. Instead, it should be taken as a crude qualitative estimate.
The possible effects of non-Gaussianity are discussed at the end of this subsection.

In the radiation-dominated epoch, the overdense region will collapse into a PBH right after the scale re-enters the horizon if the density contrast averaged over the horizon volume evaluated at the horizon re-entry $\delta(M_H)$ is larger than a threshold $\delta_{\mathrm{th}}$, where $M_H$ is the mass contained in the Hubble radius. 
The corresponding PBH mass can be evaluated as
\begin{align}
    M=\gamma M_{H}\simeq\gamma M_{\odot}\left(\frac{g_\star(T)}{10.75}\right)\left(\frac{4.2\times10^6\mathrm{Mpc}^{-1}}{k}\right)^2,
    \end{align}
where $\gamma\simeq c_s^{3}\sim0.2$ by considering a simple Jeans criterion, $g_\star(T)$ is the effective degrees of relativistic freedom at temperature $T$ at the time of horizon re-entry. 
The PBH abundance $\beta(M)$, defined as the PBH fraction of the total energy density at the time of PBH formation, can be calculated by the integration of the tail of the Gaussian distribution with $\delta>\delta_{\mathrm{th}}$,
\begin{align}
    \beta(M)=\frac{\gamma}{\sqrt{2\pi\sigma_\delta^2(H)}}\int_{\delta_{\mathrm{th}}}^{\infty}\exp\left(-\frac{\delta^2}{2\sigma_{\delta}^2(H)}\right)\mathrm{d}\delta=\gamma~\mathrm{erfc}\left(\frac{\delta_{\mathrm{th}}}{\sqrt{2}\sigma_{\delta}(H)}\right),
\end{align}
where $\sigma_{\delta}^2(H)$ is the variance of the density contrast on scale $H$, smoothed by a window function $W(k;R)$ to exclude the effect of small scale perturbations,
\begin{align}\label{sig}
\sigma_\delta^2 (M_H)=\langle \delta^2 (t, {\bf x}) \rangle =\int_0^\infty W^2(k; R) \mathcal{P}_\delta(t, k)~\mathrm{d}(\ln k),
     \end{align}
where $R=(aH)^{-1}$ and $\mathcal{P}_\delta(k)$ is the power spectrum of density contrast.
In the radiation-dominated universe, it is given by the comoving curvature perturbation spectrum as 
\begin{align}\label{Pw}
    \mathcal{P}_{\delta}(k)=\left(\frac{4}{9}\right)^2(kR)^4\mathcal{P}_{\mathcal{R}}(k)\,.
\end{align}
We adopt the Gaussian window function, $W^2(k; R)=\exp\left(-k^2R^2/2\right)$ for numerical calculation.

Taking into account the evolution of the universe after PBH formation, the PBH fraction of Cold Dark Matter (CDM) today can be expressed as \cite{Ando:2018qdb}
\begin{align}
    f_{\mathrm{PBH}}(M)\equiv\frac{\Omega_{\mathrm{PBH}}}{\Omega_{\mathrm{CDM}}}\simeq\left(\frac{\beta(M)}{1.6\times10^{-9}}\right)\left(\frac{10.75}{g_\star(T)}\right)^{1/4}\left(\frac{0.12}{\Omega_{\mathrm{DM}}h^2}\right)\left(\frac{M_{\odot}}{M}\right)^{1/2}.
\end{align}

\begin{figure}[htbp]
\centering
\includegraphics[width=0.6\textwidth]{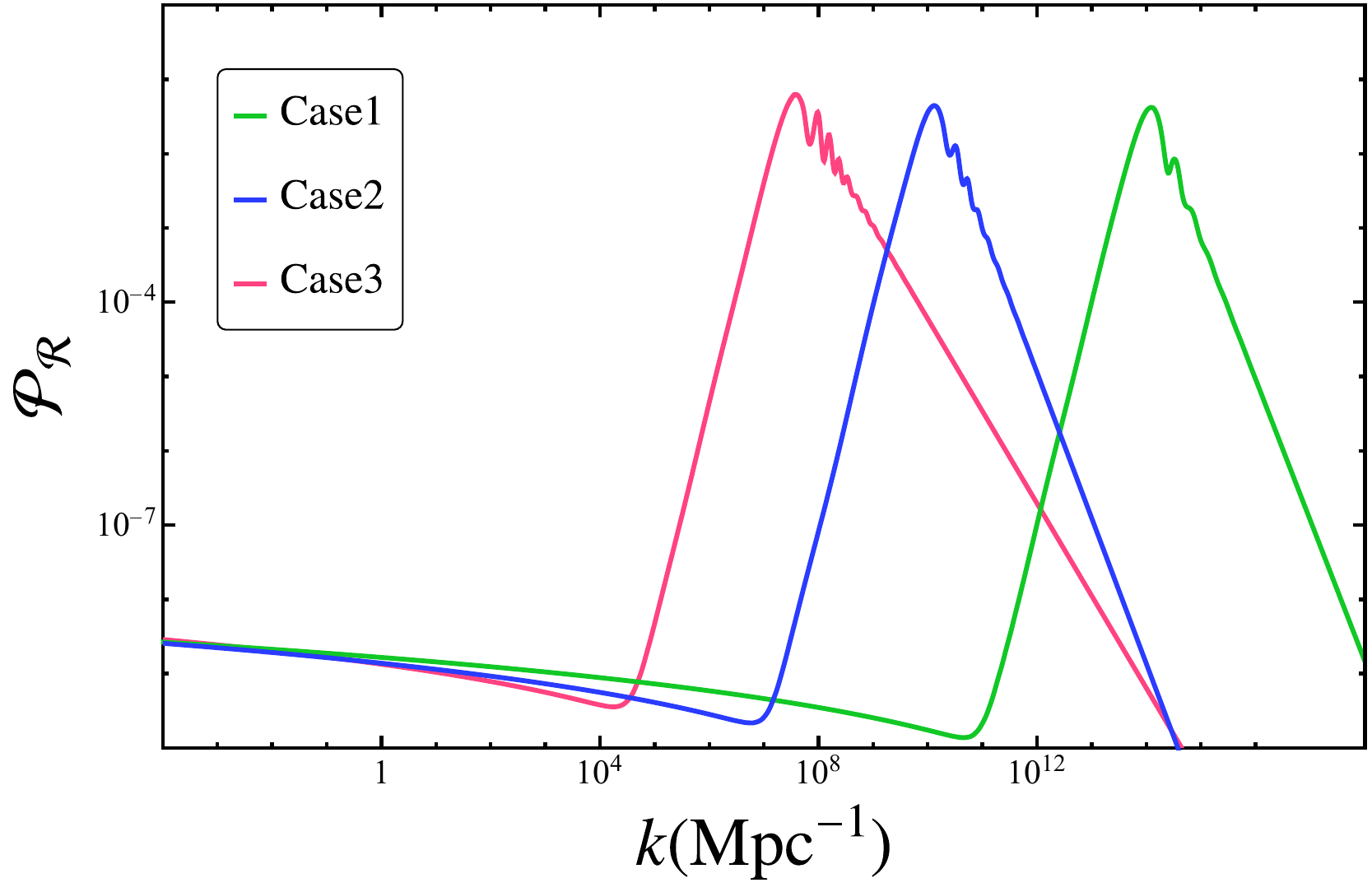}
\hspace{0.5in}
\includegraphics[width=0.6\textwidth]{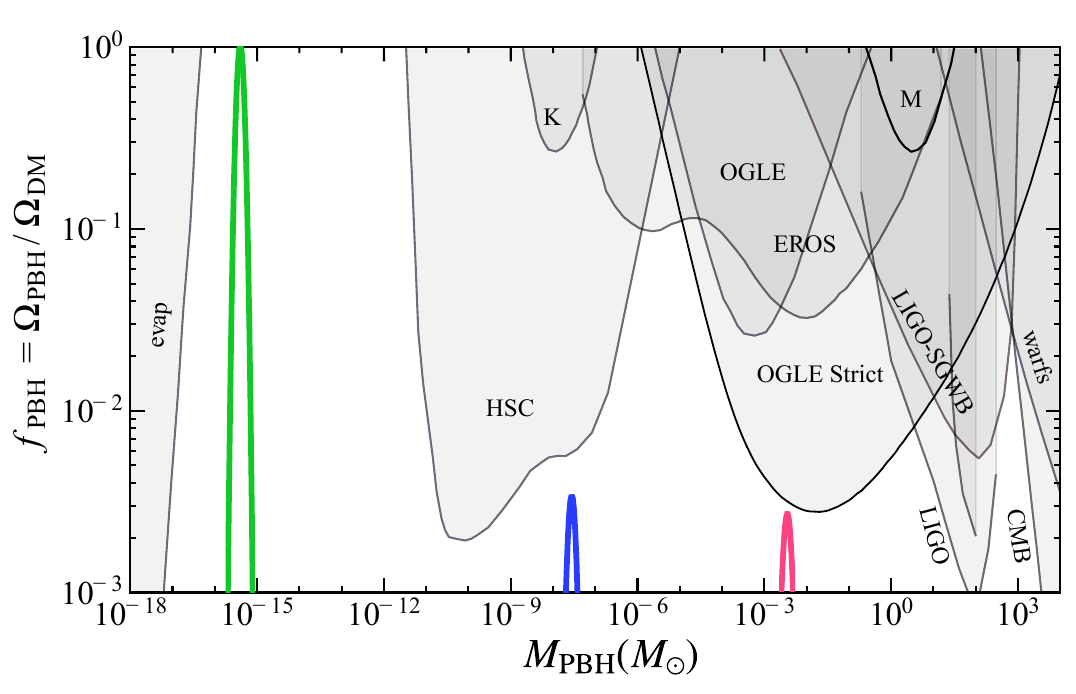}
\caption{Three numerical result of power spectrum $\mathcal{P}_{\mathcal{R}}$ and $f_\mathrm{PBH}$. Left panel: We set $\mathcal{P}_{\mathcal{R}}=2.3\times10^{-9}$ at pivot scale $k=0.002\mathrm{Mpc^{-1}}$ according to \cite{Planck:2018jri}. From right to left, the green, blue, and pink curves are numerically calculated with parameter cases 1,2,3 appearing in Table.~\ref{parameters}. We tune the parameters and obtain case 1 for PBH as a dominant candidate of cold dark matter, case 2 for the production of sub-solar PBH, and case 3 for explanation of the recently released PTAs observation intentionally. Right panel: From right to left, the green, blue, and pink curves are numerically calculated with parameter case 1,2,3 in Table.\ref{parameters}. 
The corresponding power spectrum is shown by the same color in the left panel. The $f_{\rm PBH}$ constraint plot and codes are available at \protect\hyperlink{https://github.com/bradkav/PBHbounds}{PBH bounds}, we include the observational constraints from PBH evaporation \cite{Carr:2009jm} (evap), 
HSC \cite{Croon:2020ouk}/EROS \cite{EROS-2:2006ryy}/OGLE \cite{Niikura:2019kqi}/MACHO \cite{Macho:2000nvd}/Kepler \cite{Griest:2013aaa} microlensing (HSC/EROS/OGLE/M/K) observations and millilensing observation of compact radio sources \cite{Manshanden:2018tze} (Radio), dynamical heating of ultra-faint dwarf galaxies \cite{Brandt:2016aco} (UFdwarfs), 
the accretion constraints by CMB anisotropies measured by Planck \cite{Serpico:2020ehh} (CMB), 
and background constraints by LIGO \cite{Kavanagh:2018ggo,LIGOScientific:2019kan,Chen:2019irf}
(LIGO/LIGO-SGWB). We added the latest released 20-year OGLE bounds (OGLE Strict) derived in \cite{Mroz:2024mse,Mroz:2024wag}.}
\label{fpbh}
\end{figure}

We take $\delta_{\mathrm{th}}=0.45$ and approximate $g_\star$ according to \cite{Husdal:2016haj,Franciolini:2022tfm}. The numerical results are shown in Fig.~\ref{fpbh}. The result shows that our model produces highly monochromatic $f_{\rm{PBH}}$ due to the very narrow broken power-law peak of $\mathcal{P}_{\mathcal{R}}$ at $k=k_1$. 
\begin{table}[htbp]
\centering
\begin{tabular}{cccc}
\hline
Case &1&2&3\\
\hline
     $M/M_{\mathrm{pl}}$ &$\quad 1.8\times10^{-5}\quad$&$\quad 1.59\times10^{-5}\quad$&$\quad 2.8\times10^{-5}\quad$\\
      $m/M_{\mathrm{pl}}$ &$\ 5.4\times10^{-6}\ $&$4\times10^{-6}$&$4\times10^{-6}$\\
     $\xi$ &$5/16$&$1/4$&$5/16$\\
     A&$2$&$2.3$&$1.6$\\
     $\chi_0/M$&$0.275\sqrt{1/(2\pi^2)}$&$0.172\sqrt{1/(2\pi^2)}$&$0.145\sqrt{1/(2\pi^2)}$\\
 $k_1/k_{\rm CMB}$& $6.9\times 10^{16}$& $9.4\times 10^{12}$&$3.0\times 10^{10}$\\
 $ N_*$& $40.0$& $31.7$&$26.0$\\
 $\mu^2$& $20.8$& $64.5$&$73.5$\\
  
 $\mathcal{P}_{\mathcal{R}}^{\rm peak}$& $4.16\times10^{-2}$& $4.78\times10^{-2}$&$6.21\times 10^{-2}$\\
 \hline
 \end{tabular}
\caption{The parameters used for numerical calculation.}
\label{parameters}
\end{table}
Here, $k_1$ and $ N_*\equiv n_{\star2}-n_{\rm CMB}$ are obtained from the numerical calculation, but there is a useful approximate analytical relation between $k_1$ and $k_{\rm CMB}$,
\begin{align}
    \frac{k_1}{k_{\rm CMB}}\approx \frac{H_2}{H_1}\exp(N_*)=\mu^{-1}\exp(N_*)\,,
\end{align}
where the e-folding number $N_*$ from the horizon exist epoch of the CMB scale to the end of St-2 can be approximated as
\begin{align}
     N_*\approx\frac{3}{4}(F_{\rm {CMB}}-F_{\star1})+\frac{2}{3}\ln\mu\,,
\end{align}
where $F_{\rm {CMB}}=F(\phi(n_{\rm {CMB}}))$.

\subsection{Non-Gaussianity} 

As the PBH formation occurs only at rare high peaks of the curvature perturbation, the PBH abundance is sensitive to the tail of the probability distribution function. 
Since our model produces PBHs with a highly monochromatic  $f_{\mathrm{PBH}}$, we focus on the non-Gaussianity at the comoving scale corresponding to the peak of the power spectrum. 
Although our model contains a non-trivial field metric, since the kinetic terms are essentially canonical in the sense that they are in the standard quadratic form, we expect the intrinsic scalar field perturbations to follow the Gaussian statistics. Thus the only possible non-Gaussianity is due to the non-trivial field dynamics on superhorizon scales. Such non-Gaussianities are called the local type, and they can be evaluated by expanding the $\delta N$ formula to higher orders.

We expand the comoving curvature perturbation by field perturbation $\delta\phi^I$ around $k=k_1$ using $\delta N$ formalism,\footnote{For modes $k\lesssim k_1$, $\delta N$ is dominated by the contribution of $\delta N_3$ as we show in the former section \ref{perturbations}, we neglect $\delta N_2$ because both $\phi$ and $\chi$ are heavy fields whose contributions $\delta N$ are small. So we can approximate the curvature perturbation and the non-Gaussianity by simply evaluating $\delta N_3$.}
\begin{align}
    \left.\mathcal{R}_c\right|_{k=k_1}=\delta N_3=\frac{\partial N_3}{\partial\chi}\delta\chi+\frac{1}{2}\frac{\partial^2 N_3}{\partial\chi^2}\delta\chi^2+\frac{1}{6}\frac{\partial^3 N_3}{\partial\chi^3}\delta\chi^3+\cdots,
\end{align}
where we made an approximate identification $k_1=k_2$. Using the formula for $N_3$ in \eqref{ends3}, we obtain
\begin{align}
    &\frac{\partial N_3}{\partial\chi}=-\frac{2(A-1)M_{\mathrm{pl}}^{-1}}{3\left(\sqrt{1+\frac{16}{3}\frac{(A-1)\xi}{1+1/\mu^2}}-1\right)},\\
    &\frac{\partial^2 N_3}{\partial\chi^2}=\frac{3}{2}\left(\sqrt{1+\frac{16}{3}\frac{(A-1)\xi}{1+1/\mu^2}}-1\right)\left(\frac{\partial N_3}{\partial\chi}\right)^2,\\
    &\frac{\partial^3 N_3}{\partial\chi^3}=\frac{9}{2}\left(\sqrt{1+\frac{16}{3}\frac{(A-1)\xi}{1+1/\mu^2}}-1\right)^2\left(\frac{\partial N_3}{\partial\chi}\right)^3.
\end{align}
It is customary to express the non-Gaussianities in terms of the parameters $f_{NL}$, $g_{NL}$, etc., that describe the bispectrum, trispectrum, etc. In our case we have
\begin{align}
    \mathcal{R}_c=\mathcal{R}_c^g+\frac{3}{5}f^{\rm local}_{\rm NL}(\mathcal{R}_c^g)^2+\frac{9}{25}g^{\rm local}_{\rm NL}(\mathcal{R}_c^g)^3+\cdots\,,
\end{align}
where $\mathcal{R}_c^g=({\partial N_3}/{\partial\chi})\delta\chi$ is the Gaussian part, and the non-Gaussianity parameters are given by
\begin{align}
  &f_{\mathrm{NL}}^{\mathrm{local}}=\frac{3}{10}\frac{\partial^2 N_3/\partial\chi^2}{(\partial N_3/\partial\chi)^2}=\frac{9}{20}\left(\sqrt{1+\frac{16}{3}\frac{(A-1)\xi}{1+1/\mu^2}}-1\right)\approx\frac{12}{5}\frac{(A-1)\xi}{1+1/\mu^2}\,,    \label{ng1}\\
   &g_{\mathrm{NL}}^{\mathrm{local}}=\frac{3}{50}\frac{\partial^3 N_3/\partial\chi^3}{(\partial N_3/\partial\chi)^3}=\frac43\left(f_{\mathrm{NL}}^{\mathrm{local}}\right)^2\,.\label{ng2}
\end{align}
Hence, under the assumption $(A-1)\xi\ll1$, which is the case of our model, 
the local type non-Gaussianities around the peak of the curvature perturbation spectrum are found to be small.
This agrees with the expectation that non-Gaussianities are small when the transitions are smooth, with the transition time scale being the Hubble time scale, as found in various multi-field inflationary scenarios with non-extreme transitions~\cite{Cai:2018dkf, Passaglia:2018ixg}. 

\begin{figure}[htbp]
\includegraphics[width=1\textwidth]{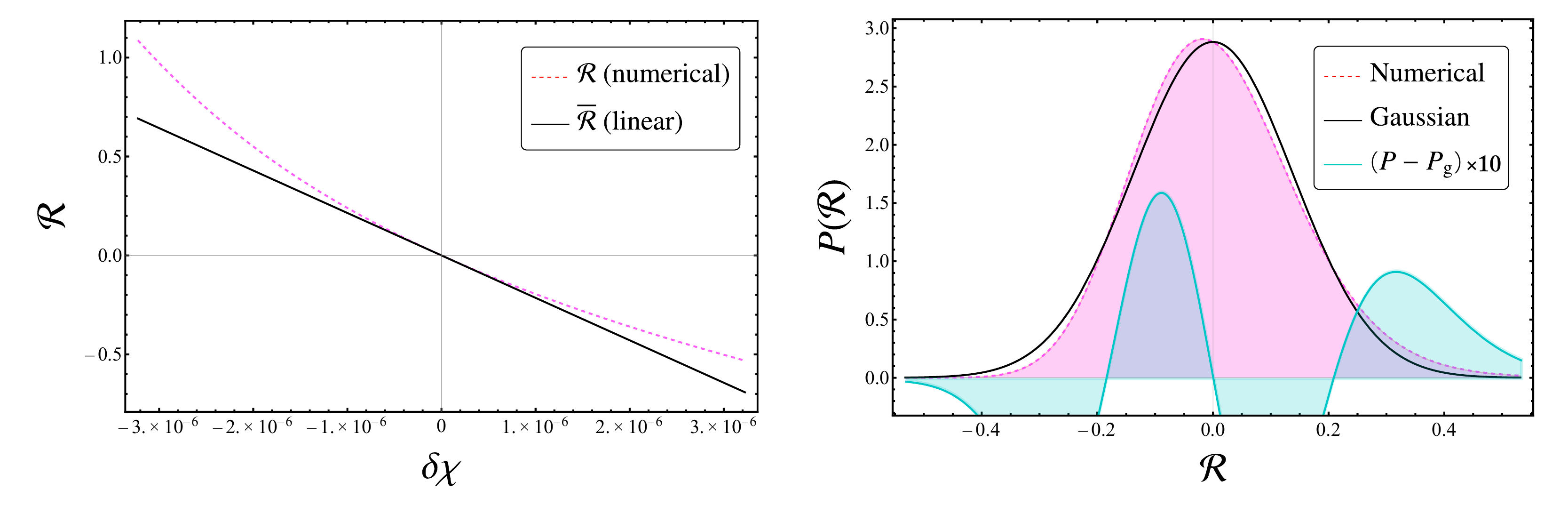}
\caption{A numerical result of local non-Gaussianity at $k=k_1$. On the left panel, the pink dotted line shows the non-linear relationship between $\mathcal{R}$ and $\delta\chi$ using numerical $\delta N$ formalism, the black solid line is given by the analytical result given in \eqref{prform}. In this figure, we take $\xi=5/16$, $A=2$. The polynomial fit of the left panel gives  $f_{\mathrm{NL}}^{\mathrm{local}}=0.285$, $g_{\mathrm{NL}}^{\mathrm{local}}=0.107$, which is consistent with \eqref{ng1} and \eqref{ng2}.}
\label{nongaussian}
\end{figure}

For the sake of completeness, we numerically computed the local non-Gaussianities around $k_1$, the results are shown by Fig.~\ref{nongaussian}. It is obvious from the left panel that the local non-Gaussianity parameters are positive and small, which are consistent with the results from \eqref{ng1} and \eqref{ng2}. 


 
\subsection{Induced GWs}
In two-stage inflationary models with an interim non-inflationary period, the power spectrum of primordial tensor perturbations deviates from the vacuum one and there may appear an oscillatory feature~\cite{Zelnikov:1991nv, Polarski:1995zn}. However, the amplitude of the oscillations is never large enough to enhance the power spectrum~\cite{Pi:2019ihn}. 
On the other hand, an enhancement of the curvature perturbation may induce tensor perturbations at second order with appreciable amplitude~\cite{Matarrese:1992rp, Matarrese:1993zf, Nakamura:1996da, Matarrese:1997ay, Noh:2004bc, Carbone:2004iv, Nakamura:2004rm, Ananda:2006af, Osano:2006ew},
and they may be detected by LIGO/Virgo/KAGRA, DECIGO, LISA, and PTA collaborations~\cite{Saito:2008jc}. 
In this subsection, we study the stochastic GW background induced at second order by the curvature perturbation. Skipping the details, the power spectrum of the induced GWs during the radiation-dominated epoch can be expressed as~\cite{Pi:2020otn, Domenech:2021ztg}
\begin{align}
\Omega_{\mathrm{GW}, \mathrm{r}}(k) & =3 \int_0^{\infty} d v \int_{|1-v|}^{1+v} d u \frac{\mathcal{T}(u, v)}{u^2 v^2} \mathcal{P}_{\mathcal{R}}(v k) \mathcal{P}_{\mathcal{R}}(u k), \\
\mathcal{T}(u, v) & =\frac{1}{4}\left[\frac{4 v^2-\left(1+v^2-u^2\right)^2}{4 u v}\right]^2\left(\frac{u^2+v^2-3}{2 u v}\right)^4\nonumber\\
&
\times\left[\left(\ln \left|\frac{3-(u+v)^2}{3-(u-v)^2}\right|-\frac{4 u v}{u^2+v^2-3}\right)^2+\pi^2 \Theta(u+v-\sqrt{3})\right].
\end{align}
Taking into account the evolution after the matter-radiation equality, the spectrum at present is given by
\begin{align}
    \Omega_{\mathrm{GW},0}(f)h^2=1.62\times10^{-5}\left(\frac{\Omega_{r,0}h^2
}{4.18\times10^{-5}}\right)\left(\frac{g_*(f)}{106.75}\right)\left(\frac{g_{*,s}(f)}{106.75}\right)^{-4/3}\Omega_{\mathrm{GW}, \mathrm{r}}(f),
\end{align}
where $\Omega_{r,0}\simeq 4.18\times10^{-5}$ is the density fraction of radiation today \cite{Planck:2018jri}, and $g_{*}$ and $g_{*,s}$ is the effective number of relativistic degrees of freedom contributing to the radiation density in the context of energy density and entropy density which are varying with temperature\cite{Husdal:2016haj} at the GW formation time, $f\approx1.5\times10^{-9}k/(1\mathrm{pc})^{-1}$Hz is the frequency corresponding to the wavenumber $k$. 

\begin{figure}[htbp]
 \centering
\includegraphics[width=0.7\textwidth]{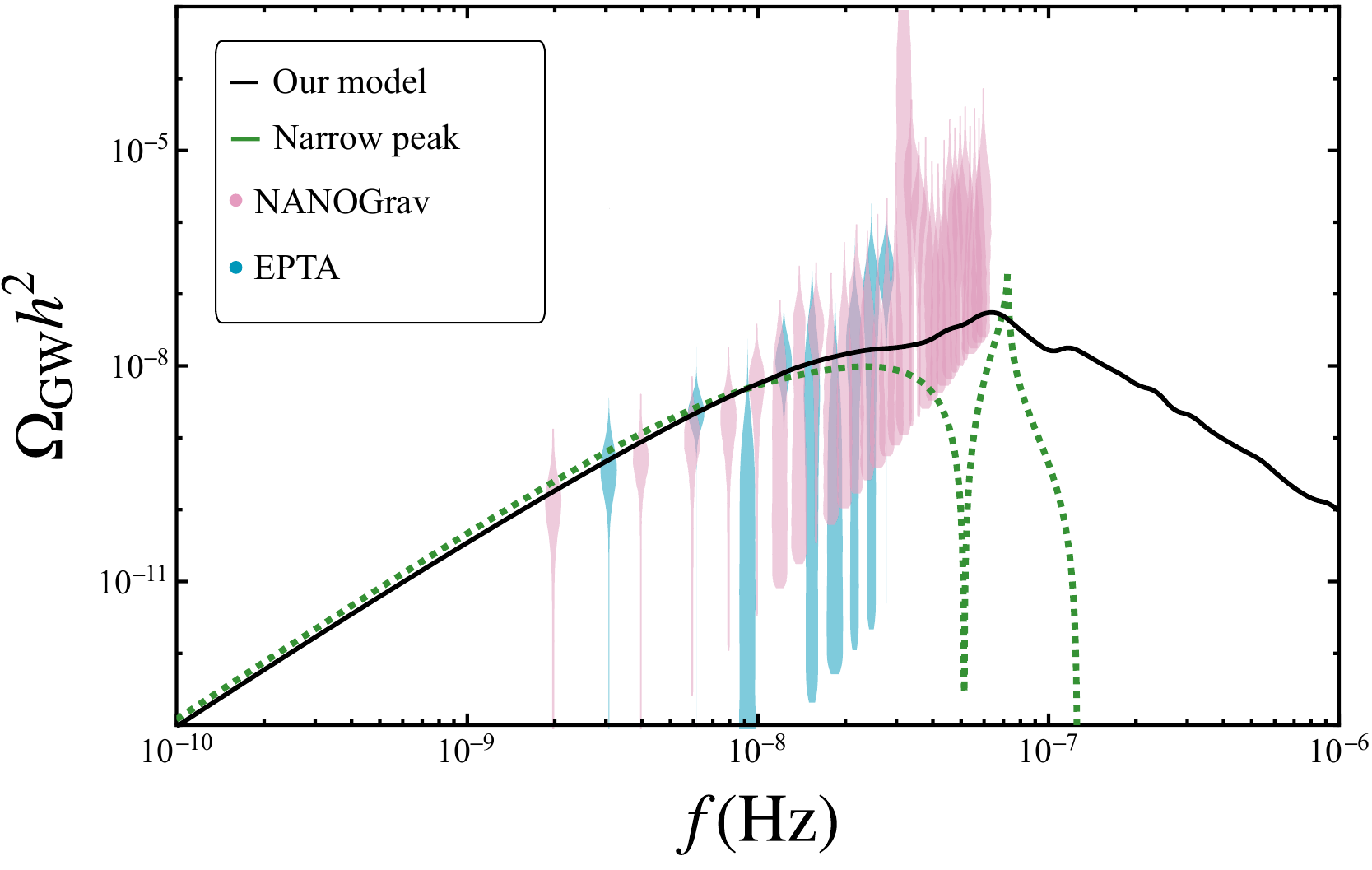}
\caption{The induced GWs produced by our model.
The parameters are those listed as case 3 in Table~\ref{parameters}. 
The violin diagrams are from NANOGrav-15yr~\cite{NANOGrav:2023hvm} (pink) and EPTA (light blue)~\cite{EPTA:2023xxk}. 
The dotted green curve is the analytical narrow peak spectrum given by Eq.~\eqref{narrowpeak}, with the parameters $\mathcal{A}_{\mathcal{R}}=0.025$ and $\Delta=0.05$.}
\label{fitng}
\end{figure}

As an example, we adopt the parameters of case 3 in Table~\ref{parameters} calculate the corresponding GW power spectrum, and compare it with the recent PTA data. 
There appears infrared behavior of $\sim f^{3}$ at the far-IR end and $\sim f^{2.5}$ at the near-IR end. These infrared features are consistent with the GWs induced by a lognormal spectrum, 
\begin{equation}
\mathcal{P}_{\mathcal{R}}(k)=\frac{\mathcal{A}_{\mathcal{R}}}{\sqrt{2\pi}\Delta}
\exp \left[-\frac{\ln^2\left(k/ k_\mathrm{peak}\right)}{2\Delta^2}\right]\,.
\end{equation}
The resulting GW power spectrum can be approximated by \cite{Pi:2020otn}
\begin{align}\label{narrowpeak}
&\Omega_{\mathrm{GW}, \mathrm{r}} \approx  3 \mathcal{A}_{\mathcal{R}}^2 \kappa^2 e^{\Delta^2}\left[\operatorname{erf}\left(\frac{1}{\Delta} \operatorname{arcsinh} \frac{\kappa e^{\Delta^2}}{2}\right)-\operatorname{erf}\left(\frac{1}{\Delta} \operatorname{Re}\left(\operatorname{arccosh} \frac{\kappa e^{\Delta^2}}{2}\right)\right)\right]\nonumber\\
&\times\left(1-\frac{1}{4} \kappa^2 e^{2 \Delta^2}\right)^2
\left(1-\frac{3}{2} \kappa^2 e^{2 \Delta^2}\right)^2\nonumber\\
&\times\left\{\left[\frac{1}{2}\left(1-\frac{3}{2} \kappa^2 e^{2 \Delta^2}\right) \ln \left|1-\frac{4}{3 \kappa^2 e^{2 \Delta^2}}\right|-1\right]^2+\frac{\pi^2}{4}\left(1-\frac{3}{2} \kappa^2 e^{2 \Delta^2}\right)^2 \Theta\left(2-\sqrt{3} \kappa e^{\Delta^2}\right)\right\},
\end{align}
for $\Delta\ll1$ with $\kappa=k/k_\mathrm{peak}$. 
In Fig.~\ref{fitng} we plot the GW power spectrum. The black solid curve corresponds to our model with parameters fixed in case 3 of Table~\ref{parameters}. 
The pink and blue violin diagrams are observational data constrained by NANOGrav-15yr ~\cite{NANOGrav:2023hvm} and EPTA~\cite{EPTA:2023xxk} collaborations, respectively. 
It is intriguing to see that the power spectrum of our model fits the current observational data within the error bar. We, however, caution that we have not done any statistical analysis about the goodness of fit.
As a reference, we also plot the spectrum with a lognormal peak formulated in \eqref{narrowpeak} by the green dashed curve which shares the same infrared behavior $k^3$ on low frequency which is the
universal infrared scaling of GW spectrum~\cite{Cai:2019cdl}.

\section{Conclusion and Discussion}
\label{conclusion}
In this paper, we investigated the two-field two-stage inflationary scenario having an interim non-inflationary period by adopting Starobinsky's $R^2$-gravity endowed with a non-minimally coupled scalar field $\chi$ (illustrated in Fig.~\ref{pertevo}). 
Upon a conformal transformation, this model is equivalent to a biscalar-tensor theory in the Einstein frame, with the scalaron $\phi$ arising from the $R^2$ term and the non-minimal scalar $\chi$ being the two scalar fields.
The first stage is dominated by $\phi$. Namely, $\phi$ is the light field that undergoes slow-rolling, while $\chi$ is heavy. We call it St-1.
After the end of St-1, $\phi$ undergoes damped oscillations, which breaks inflation. We call this stage St-2. If the potential were zero at the minimum of oscillations, inflation would have ended.
However, the non-vanishing potential at the minimum leads to the second stage of inflation, now dominated by $\chi$ which becomes light while $\phi$ becomes heavy and stays at its local minimum.
We call this second inflationary stage St-3.

We assume the end of St-1 to be much later than the time when the comoving wavenumber corresponding to the CMB scale, $k_{\rm CMB}$, crosses the horizon radius, say 30-40 e-folds after the horizon exit of the CMB scale. We denote the corresponding comoving scale by $k_1$.
One important feature of our model is that toward the end of St-1, isocurvature perturbations due to $\delta\chi$ start to grow rapidly and give rise to a significant enhancement of the curvature perturbation at scales $k\sim k_1$.
We solved the equations of motion for the background and linear perturbations both analytically and numerically. Using the $\delta N$ formalism, it was found that within a wide range of model parameters, the enhancement factor can reach $\sim10^4$ (or $\sim10^8$ in terms of the power spectrum). 
This means the RMS amplitude of the curvature perturbation at $k=k_1$ may exceed $10^{-1}$, considering that it is $\sim10^{-5}$ on the CMB scale, As a consequence, PBHs may form from rare peaks of the curvature perturbation whose amplitude exceed a threshold value of $O(1)$ when the scale re-enters the horizon during the radiation-dominated epoch. 

In terms of the $\delta N$ formalism, this enhancement may be understood as a consequence of the sharp turn of the inflationary trajectory in the field space, an almost orthogonal turn from the $\phi$-direction to the $\chi$-direction. This converts the original isocurvature perturbation due to $\delta\chi$ at St-1 to an adiabatic perturbation at St-3. Since $\delta\chi$ decays like $\sim a^{-3/2}$ during St-1, the isocurvature effect is negligible at early stages. 
Conversely, it means that the isocurvature effect grows like $a^{3/2}$ toward the end of St-1. 
This leads to the growth of the curvature perturbation spectrum proportional to $k^3$. This is our new finding. This growth rate is universal as long as the isocurvature field ($\chi$ in our case) is heavy enough at the first inflationary stage and dominates the second inflationary stage as a slow-rolling inflaton. 

If the isocurvature field is not heavy enough, it may also undergo over-damping at the first stage. In such a case, one expects the growth rate to be smaller than $k^3$. In this sense, the growth $\propto k^3$ is the maximum growth rate one can obtain in the standard two-field, two-stage models with a sharp turn of the inflationary trajectory in the field space.

Due to the steep enhancement feature in the power spectrum, there appears a sharp peak in the PBH mass spectrum. This implies that PBH masses in our model will be almost monochromatic. The PBH mass $M_\mathrm{PBH}$ is determined by the wavenumber of the peak which differs according to the different values of parameters of the model.

In addition to the PBH formation, we also discussed the GW background induced by the curvature perturbation at second order. The enhanced curvature perturbation can give rise to a GW spectrum large enough to be of observational interest~\cite{Saito:2008jc}. 
In particular, we computed the induced GWs corresponding to the PBH mass of $M_\mathrm{PBH}\sim10^{2}M_\odot$ and found that they fit the NANOGrav-15yr data~\cite{NANOGrav:2023hvm} and EPTA data~\cite{EPTA:2023xxk} reasonably well.

In this paper, we assumed that the curvature perturbation is Gaussian. However, it is known that non-Gaussianities may have a significant influence on the PBH abundance~\cite{Young:2013oia, Yoo:2019pma, Atal:2019cdz, Ferrante:2022mui, Ianniccari:2024bkh,Pi:2022ysn} as well as on the PBH clustering~\cite{Desjacques:2018wuu, Suyama:2019cst, Franciolini:2018vbk}, which may also leave characteristic features on the corresponding induced GWs~\cite{Cai:2018dig, Garcia-Bellido:2016dkw, Nakama:2016gzw, Unal:2018yaa, Adshead:2021hnm, Garcia-Saenz:2022tzu,Liu:2023ymk, Cai:2018dig}. 
In our model, we found that the non-Gaussianities are of local type, and those generated around the peak of the power spectrum are very small, as expected in the case of multi-field inflationary scenarios with non-extreme transitions~\cite{Cai:2018dkf, Passaglia:2018ixg}.

Our model gives a prospective aspect of multi-stage inflation with each stage effectively dominated by a single slow-roll field. At every turn of the trajectory from one field direction to the other in the field space, there may be a significant enhancement of the curvature perturbation.
Depending on how the fields are coupled to each other, one may find distinct features in the perturbation spectrum that may be testable by observation. In this paper, we only considered a short non-inflationary interim stage. It may be interesting to consider a model with a prolonged non-inflationary interim stage. If the energy scale of the final inflationary stage is very low, it may have interesting implications for particle physics phenomenology.
We leave all these issues for future work.

\section{Acknowledgement}
We thank Shi Pi, Jinsu Kim, and Minxi He for useful discussions. 
This work is supported in part by JSPS KAKENHI Grant Nos. JP20H05853 and JP24K00624.
XW is supported by China Scholarship Council No. 202306260140,
YZ is supported by the Fundamental Research Funds for the Central Universities,  and by Project 12047503 supported by NSFC, and by China Scholarship Council. 
XW and YZ gratefully acknowledge the hospitality and support of the Kavli Institute of Physics and Mathematics of the Universe (Kavli IPMU), the University of Tokyo.

\appendix
\section{A general description of the field equations}
We can introduce an induced metric of the field space $h_{ab}$ to rewrite this system in a compact way \cite{Sasaki:1995aw,GrootNibbelink:2000vx,Achucarro:2010da,He:2018gyf}  $\psi^a=(\phi,\chi)$,
\begin{align}
    h_{ab}=\begin{pmatrix}
    1\quad 0\\
    0\quad F^{-1}
    \end{pmatrix},\quad a,b=1,2.
\end{align}
We can define the Christoffel symbol and the Riemann curvature tensor for the curved field space as
\begin{align}
    \Gamma^{a}_{bc}&\equiv\frac{1}{2}h^{ad}(\partial_bh_{cd}+\partial_ch_{db}-\partial_{d}h_{bc}),\\
    R^{a}_{bcd}&\equiv\partial_c\Gamma^a_{db}-\partial_d\Gamma^a_{cb}+\Gamma^a_{be}\Gamma^e_{dc}-\Gamma^a_{de}\Gamma^e_{cb}.\label{Rfield}  
\end{align}
The non-vanishing terms in our model are the following:
\begin{equation}
\begin{aligned}\label{fieldcompo}
    &\Gamma^{1}_{22}=\sqrt{\frac{1}{6}}\frac{1}{M_\mathrm{pl}}F^{-1},\qquad
    \Gamma^{2}_{12}=-\sqrt{\frac{1}{6}}\frac{1}{M_\mathrm{pl}},\qquad
    \Gamma^{2}_{21}=\Gamma^{2}_{12},\\  
    &R^{1}_{212}=-\frac{1}{6}\frac{1}{M_\mathrm{pl}^2}F^{-1},\qquad
    R^{2}_{112}=\frac{1}{6}\frac{1}{M_\mathrm{pl}^2},\qquad
    R^{1}_{221}=\frac{1}{6}F^{-1}\frac{1}{M_\mathrm{pl}^2},\qquad
    R^{2}_{121}=-\frac{1}{6}\frac{1}{M_\mathrm{pl}^2}.
\end{aligned}  
\end{equation}
Taking the spatially flat FLRW metric $\tilde{g}_{\mu\nu}=\mathrm{diag}(-1,a^2,a^2,a^2)$ as the background space time metric, the background equations of motion for $\psi^a$ can be expressed as
\begin{align}
    \frac{D\dot\psi^a}{dt}+3H\dot\psi^a+h^{ab}U_{,b}=0,
\end{align}
while the components of Einstein field equations are
\begin{align}
    &3M_\mathrm{pl}^2H^2=\frac{1}{2}h_{ab}\dot\psi^a\dot\psi^b+U(\psi),\\
    &2M_\mathrm{pl}^2\dot H=-h_{ab}\dot\psi^a\dot\psi^b,
\end{align}
where $\dot\psi^a\equiv d\psi^a/dt$, while ${D\dot\psi^a}/{dt}\equiv\frac{d\dot\psi^a}{dt}+\Gamma^a_{bc}\dot\psi^b\dot\psi^c$ is the covariant derivative in the curved field space. 

For a perturbed space-time, the linear scalar perturbations are expressed as
\begin{align}
    ds^2=-(1+2A)dt^2+2a\partial_i Bdx^idt+a^2[(1-2\mathcal{R})\delta_{ij}+2\partial_i\partial_j E]dx^idx^j,
\end{align}
where $\mathcal R$ is curvature perturbation.
Defining the gauge invariant variables,
\begin{align}
    Q^a\equiv\delta\psi^a+\frac{\dot\psi^a}{H}\mathcal{R}\,,
\end{align}
which correspond to the field perturbations on flat slicing, 
the equations of motion are expressed as~\cite{Sasaki:1995aw}
\begin{align}
    \frac{D^2Q_k^a}{dt}+3H\frac{DQ_k^a}{dt}+\frac{k^2}{a^2}{Q_k^a}+U^{;a}_{\ \ ;b}Q_k^b-R^a{}_{bcd}\dot\psi^b\dot\psi^cQ_k^d-\left[\frac{1}{a^3}\frac{d}{dt}\left(\frac{a^3}{H}\dot\psi^a\dot\psi^b\right)\right]h_{bc}Q^b=0,
    \label{perteqQ}
\end{align}
where $k$ is the comoving wave number, $U$ is the potential. The explicit forms of the potential derivatives are 
\begin{align}
   U^{;\phi}{}_{;\phi}=&\frac{M^2 \left(\xi  (\chi -\chi_0)^2M_{\mathrm{pl}}^{-2}-1\right)}{F}
   \nonumber\\
   &+\frac{2 \left(\chi ^2+3 M^2 \left(\xi  (\chi -\chi_0)^2M_{\mathrm{pl}}^{-2}-1\right)^2+4 V(\chi)M_{\mathrm{pl}}^{-2}\right)}{3 F^2}\,,\\
   U^{;\chi}{}_{;\chi}=&3 M^2 \xi+\frac{6 \lambda  \chi ^2-2 m^2+M^2 \left(\xi  \left((18 \xi +1) (\chi -\chi_0)^2M_{\mathrm{pl}}^{-2}-6\right)-1\right)}{2 F}
   \notag\\&+\frac{3 M^2 \left(\xi  (\chi -\chi_0)^2M_{\mathrm{pl}}^{-2}-1\right)^2+4 V(\chi)M_{\mathrm{pl}}^{-2}}{6 F^2}\,,\\
 U^{;\phi}{}_{;\chi}=&-\sqrt{\frac{3}{2}}\frac{ M^2 \xi  (\chi -\chi_0)M_{\mathrm{pl}}^{-1}}{F}
 \nonumber\\
 &+\sqrt{\frac{3}{2}} \frac{\left(-\lambda  \chi ^3+m^2 \chi -3 M^2 \xi  (\chi -\chi_0) \left(\xi  (\chi -\chi_0)^2M_{\mathrm{pl}}^{-2}-1\right)\right)M_{\mathrm{pl}}^{-1}}{F^2}\,,\\
    U^{;\chi}{}_{;\phi}=&FU^{;\phi}{}_{;\chi}.
\end{align}

During the first stage, St-1, since $\chi\approx\chi_0$, the radio of off-diagonal to the diagonal terms are
\begin{equation}
    \begin{aligned}
\left|\frac{{U}^{;\phi}{}_{;\chi}}{U^{;\phi}{}_{;\phi}}\right|&
\simeq \sqrt{\frac{3}{2}}\frac{ m^2 \chi_0}{F M^2M_{\mathrm{pl}}}\ll 1,\\
\left|\frac{U^{;\phi}{}_{;\chi}}{U^{;\chi}{}_{;\chi}}\right|
&\simeq\frac{m^2 \chi_0}{\sqrt{6} F^2 M^2 \xi M_{\mathrm{pl}}}\ll 1,
\end{aligned}
\label{offdiag1}
\end{equation}
under the approximations $\mu^2\gg1$ and $F\gg 1$.
During the third stage, St-3, we have $F\approx1$. Hence for $\chi_0<\chi\ll\chi_g$,
\begin{equation}
    \begin{aligned}   
   &\sqrt{\frac{3}{2}}\frac{ m^2 \chi_0}{M^2M_{\mathrm{pl}}}<\left|\frac{U^{;\phi}{}_{;\chi}}{U^{;\phi}{}_{;\phi}}\right|\ll\sqrt{\frac{3}{2}}\frac{ (2 A-3) \sqrt{A \xi }}{(A-1)(A-2)}\,,\\
    &\frac{m^2 \chi_0}{\sqrt{6} M^2 \xi M_{\mathrm{pl}}}<\left|\frac{U^{\phi}{}_{;\chi}}{U^{;\chi}{}_{;\chi}}\right|\ll\frac{\sqrt{6} (2 A-3)  \sqrt{A \xi }}{4 A^2 (m/M)^2+A (18 \xi -1)+1}\,.\notag\\
\end{aligned}
   \label{offdiag2}
\end{equation}
Thus for the two slow-roll stages, we may neglect the off-diagonal parts (that is, the coupling between $\chi$ and $\phi$) when solving the perturbation equations of motion. 
Then $\epsilon^a_b$ defined in \eqref{epsilonab} are given by
\begin{align}
    &\epsilon^{\phi}_{\phi}=-\frac{4 \left(3 M^2M_{\mathrm{pl}}^{2} \left(\xi  (\chi -\chi_0)^2M_{\mathrm{pl}}^{-2}-1\right) \left(F+2 \left(\xi  (\chi -\chi_0)^2M_{\mathrm{pl}}^{-2}-1\right)\right)+8V(\chi)\right)}{3 \left(3 M^2M_{\mathrm{pl}}^{2} \left(F+\xi  (\chi -\chi_0)^2M_{\mathrm{pl}}^{-2}-1\right)^2 \left(1-6 \lambda  \left(F+\xi  (\chi -\chi_0)^2M_{\mathrm{pl}}^{-2}-1\right)\right)+4V(\chi)\right)}\,,\\
    &\epsilon^{\chi}_{\chi}=-\frac{2 \left(18 F^2 M^2 \xi +3 F \left(6 \lambda  \chi ^2-2 m^2+M^2 \left(\xi  \left((18 \xi +1) (\chi -\chi_0)^2M_{\mathrm{pl}}^{-2}-6\right)-1\right)\right)\right)}{3 \left(3 M^2 \left(F+\xi  (\chi -\chi_0)^2M_{\mathrm{pl}}^{-2}-1\right)^2 \left(1-6 \lambda  \left(F+\xi  (\chi -\chi_0)^2M_{\mathrm{pl}}^{-2}-1\right)\right)+4V(\chi)M_{\mathrm{pl}}^{-2}\right)}\notag\\
    &-\frac{2\left(3 M^2 \left(\xi  (\chi -\chi_0)^2M_{\mathrm{pl}}^{-2}-1\right)^2+4 V(\chi)M_{\mathrm{pl}}^{-2}\right)}{3 \left(3 M^2 \left(F+\xi  (\chi -\chi_0)^2M_{\mathrm{pl}}^{-2}-1\right)^2 \left(1-6 \lambda  \left(F+\xi  (\chi -\chi_0)^2M_{\mathrm{pl}}^{-2}-1\right)\right)+4V(\chi)M_{\mathrm{pl}}^{-2}\right)}\,,
\end{align}
where we have made an approximation $\epsilon_H=0$.

\section{The comoving curvature and isocurvature perturbations}\label{comovingper}
In this appendix, following \cite{Gordon:2000hv,He:2018gyf}, we show the explicit expressions for the comoving curvature perturbation and the isocurvature perturbation in our model.

The comoving slicing can be considered for each field, on which the field value is homogeneous. The curvature perturbation on the comoving slices for the field $\psi^a$ is defined as
\begin{align}
    \mathcal{R}_c^a\equiv\mathcal{R}+\frac{H}{\dot{\psi}^{a}}\delta\psi^a=\frac{H}{\dot{\psi}^{a}}Q^a,
\end{align}
where $\mathcal{R}$ and $\delta\psi^a$ are the curvature perturbation and the field fluctuations evaluated on an arbitrary slicing, and $Q^a$ are the field fluctuations on the flat slicing $\mathcal{R}=0$.
Then, the total comoving curvature perturbation is defined as
\begin{align}
\mathcal{R}_c\equiv\mathcal{R}+H\frac{h_{ab}\dot\psi^a\delta\psi^b}{\dot\psi_0^2}\,,  
\label{Rcorigin}
\end{align}
where 
\begin{align}
\dot\psi_{0}^2\equiv
h_{ab}\dot\psi^a\dot\psi^b=\dot\phi^2+F^{-1}\dot\chi^2.
\end{align}
 
To evaluate the comoving curvature perturbation and the isocurvature perturbation, we work in the flat slicing $\mathcal{R}=0$. The field perturbation can be decomposed into the components parallel and orthogonal to the background trajectory, 
\begin{align}
    Q_{\parallel}&\equiv h_{ab}T^aQ^b=\frac{\dot\phi\delta\phi+F^{-1}\dot\chi\delta\chi}{\sqrt{\dot\phi^2+F^{-1}\dot\chi^2}},\label{Q1}\\
    Q_{\perp}&\equiv h_{ab}N^aQ^b\,,\label{Q2}
\end{align}
where $T^a$ and $N^a$ are the unit vectors parallel and normal to the trajectory, respectively, given by
\begin{align}
    T^a&\equiv\frac{\dot\psi^a}{|\dot\psi_0|}=\frac{1}{\sqrt{\dot\phi^2+F^{-1}\dot\chi^2}}\left(\dot\phi,\dot\chi\right),\\
    N^a&\equiv\frac{D{T^a}/dt}{\left|D{ T^a}/dt\right|}\,; \quad \frac{DT^a}{dt}=\frac{dT^a}{dt}+\Gamma^a_{bc}\dot\psi^bT^c\,.
    \label{Nadef}
\end{align}
In our model, we have
\begin{align}\label{dT1}
    \frac{DT^1}{dt}
    &=
  \frac{1}{\dot\psi_0^3}F^{-1}\dot\chi\left({\frac{D\dot\phi}{dt}\dot\chi-\frac{D\dot\chi}{dt}\dot\phi}\right)\,,
\end{align}
and similarly
\begin{align}\label{dT2}
    \frac{DT^2}{dt}=\frac{1}{\dot\psi_0^3}F^{-1/2}\dot\phi\left(\frac{D\dot\chi}{dt}\dot\phi-\frac{D\dot\phi}{dt}\dot\chi\right).
\end{align}
Inserting \eqref{dT1} and \eqref{dT2} into \eqref{Nadef}, we obtain
\begin{align}\label{Naexp}
    N^{a}=\frac{F^{1/2}}{\dot\psi_{0}}(F^{-1}\dot\chi,-\dot\phi)\,.
\end{align}

The curvature perturbation on the comoving slice $\mathcal{R}_{c}$ and the isocurvature perturbation $\mathcal{S
}$ are given by
\begin{align}
    \mathcal{R}_{c}&=\frac{H}{\dot\psi_{0}}Q_{\parallel}\,,
    \label{calRcbyQ}\\
    \mathcal{S}&=\frac{H}{\dot\psi_{0}}Q_{\perp}\,.
\end{align}
Note that \eqref{calRcbyQ} is the same as \eqref{Rcorigin}.
Inserting \eqref{Naexp} into \eqref{Q2}, together with \eqref{Q1}, we obtain the explicit expressions for $\mathcal{R}_c$ and $\mathcal{S}$ as
\begin{align}
\mathcal{R}_c&=H\frac{\dot\phi\delta\phi+F^{-1}\dot\chi\delta\chi}{\dot\phi^2+F^{-1}\dot\chi^2}\,,\label{Rcfin}\\
\mathcal{S}&=HF^{-1/2}\frac{\dot\chi\delta\phi-\dot\phi\delta\chi}{\dot\phi^2+F^{-1}\dot\chi^2}\,.\label{Sfin}
\end{align}

\section{Parameters}
Here we summarize our model parameters. We have 5 independent parameters as shown in Table.~\ref{parameters}, namely, $M$, $m$, $\xi$, $\chi_0$ and $A$ (or $\mu$). Below are their useful relations to the parameters of the potential,
\begin{equation}
    \begin{aligned}
    &\mu^2=\frac{3M^2M_{\rm pl}^2}{4V_0}=\frac{3A\xi M^2}{m^2},\\
    &\lambda=A\xi \frac{m^2}{M_{\mathrm{pl}}^2},\\
    &V_0=\frac{m^4}{4\lambda}=\frac{m^2M_{pl}^2}{4A\xi},\\
    &\chi_g^2=\frac{M^2_{\mathrm{pl}}}{A\xi},\\
\end{aligned}
\end{equation}
We summarize the constraints that the parameters should satisfy for our model to be viable in Table.~\ref{parameterres}.

\begin{table}[htbp]
\centering
\begin{tabular}{cp{8cm}}
\hline
Constraints & Conditions\\
\hline
     $\mu^2\gtrsim3+2\sqrt{3}$&For the existence of the interim non-inflationary stage (see \eqref{phistar}).\\
      $1<A\leq\frac{\sqrt{3}}{3}\mu^2\left(1+\sqrt{1-(3+2\sqrt{3})/\mu^2}\right)$&For the transition of $M_{\chi,\rm eff}^2$ from positive to negative during St-2 (see \eqref{lambdaupper}).\\
     $\mathcal{O}(0.1)M\sqrt{1/(2\pi^2)}\lesssim \chi_0\ll\chi_g$& To keep $\delta\chi<\chi$ when $\chi$ dominates the power spectrum.\\
     $\xi(A-1)\ll1$& To satisfy the slow roll condition during St-3 (see \eqref{xiAapp}).\\
   \hline
 \end{tabular}

\caption{The constraints to the parameters.}
\label{parameterres}
\end{table}

\bibliographystyle{utphys}
\bibliography{citation}
\end{document}